\documentclass[reprint,twocolumn,superscriptaddress,longbibliography]{revtex4-2}
\usepackage{graphicx,color}
\usepackage{amsmath}
\usepackage{amsthm}
\usepackage{amssymb}
\usepackage{physics}
\usepackage{amsfonts}
\usepackage{tabularx}
\usepackage{comment}
\usepackage[normalem]{ulem}
\usepackage{soul}
\UseRawInputEncoding

\newcommand{\be}{\begin{equation}} 
\newcommand{\ee}{\end{equation}} 
\newcommand{\bea}{\begin{eqnarray}} 
\newcommand{\eea}{\end{eqnarray}} 
 
\newcommand{\lam}{\lambda}

\newcommand{\zet}{\zeta}

\newcommand{\meq}{\mu_{\text{eq}}}

\begin{document}
\title{Critical scaling and supercritical coarsening in Active Model B$+$}
\author{Abir Bhowmick}
\thanks{ab23rs019@iiserkol.ac.in}
\affiliation {Department of Physical Sciences, Indian Institute of Science Education and Research Kolkata, Mohanpur, 741246, India.}
\author{P. K. Mohanty}
\thanks{pkmohanty@iiserkol.ac.in}
\affiliation {Department of Physical Sciences, Indian Institute of Science Education and Research Kolkata, Mohanpur, 741246, India.}
\affiliation{Max-Planck Institute for Physics of Complex Systems,  Nöthnitzer Str. 38, 01187 Dresden, Germany.}

\begin{abstract}
We study critical dynamics and phase-ordering kinetics in Active Model B (AMB) and its minimal extension, Active Model B$+$ (AMB$+$), using deterministic simulations in two dimensions. At criticality $r_c=0$, both models display identical mean-field scaling despite nonequilibrium currents, with order-parameter decay with time as $m(t)\sim t^{-\alpha}$, with $\alpha=\frac14$, and dynamical exponent being $z=4$. A generalized equal-area construction yields the binodal densities and phase diagram of AMB$+$. For supercritical quenches, domain size grows as $L(t)\sim t^{1/3}(1+c/\ln t)$, revealing logarithmic corrections to the classic  $t^{1/3}$ growth-law; moreover it is consistent with  the functional renormalization group predictions for marginal activity in $d=2$. 
While the logarithmic corrections are quite prominent in AMB, in AMB$+$ they are suppressed as the active current acts against the formation of macro-clusters; the growth is eventually arrested when a long-lived microphase-separated state appears.
\end{abstract}

\maketitle

\section{Introduction}
Active matters are nonequilibrium systems in which individual agents continuously consume energy from the environment to self-propel~\cite{Ramaswamy2017, Marchetti2013}. Examples span a wide range of scales, including schools of fish~\cite{Katz2011}, flocks of birds~\cite{Ballerini2008}, active colloidal suspensions~\cite{Marchetti2016, Farage2015, Liebchen2017, Illien2017, vanderLinden2019, Zttl2023, Melio2025}, the cytoskeleton of cells~\cite{Bendix2008, Sanchez2012, Alvarado2013, Geisterfer2021}, molecular motors~\cite{Jlicher1997, Ndlec1997, Kruse2004, Liverpool2003, Liverpool2008}, and actin filaments~\cite{Schaller2010}. Because energy is injected locally at the microscopic scale, these systems generically violate detailed balance, allowing additional symmetry-permitted terms in their coarse-grained dynamics. The resulting local breaking of time-reversal symmetry leads to novel collective phenomena absent in equilibrium systems. A prominent example of these kind of systems, is motility-induced phase separation (MIPS), in which self-propelled particles spontaneously phase separates into dense and dilute phases due to density-dependent motility~\cite{Tailleur2008,Cates2015,Fily2012}. 
This mechanism has been extensively confirmed in simulations~\cite{Fily2012,Redner2013,Stenhammar2013,Stenhammar2014,Bhowmick2025} and experiments~\cite{Buttinoni2013,Liu2019}, and has motivated the development of continuum field theories describing phase separation in active matter systems~\cite{Cates2015, Shaebani2020}. 

Motivated by the apparent similarity between MIPS and equilibrium liquid--vapor phase separation, early theoretical efforts sought to extend equilibrium phase-separation formalisms to active systems\cite{Tailleur2008,Cates2015,Solon2015}, 
speculating that time-reversal symmetry (TRS) is restored  in the steady state at the macroscopic level~\cite{Farage2015, Speck2014, Nardini2017}.
The equilibrium liquid--vapor phase separation with a conserved scalar order parameter $\phi$ is described by the \textit{Cahn--Hilliard} equation~\cite{Cahn1958}, or also known as Model~B in the classification of \textit{Hohenberg} and \textit{Halperin}~\cite{Hohenberg1977}. Here, the dynamics obey detailed balance, and the chemical potential derives from a free-energy functional, $\meq=\delta\mathcal{F}/\delta\phi$. Consequently, both critical behavior and late-stage coarsening are well understood: the upper critical dimension is $d_c=4$, static critical exponents belong to the Ising universality class, and domain growth after a deep quench follows the Lifshitz--Slyozov law $L(t)\sim t^{1/3}$ for $d\ge2$~\cite{Bray1994}. Importantly, within the renormalization-group framework, equilibrium Model B possesses no marginal operators in $d=2$, so logarithmic corrections to scaling are not expected.

Despite these equilibrium-inspired descriptions of MIPS, accumulating evidence indicates that active phase separation cannot, in general, be reduced to an effective equilibrium process. Subsequent numerical investigations of repulsive active Brownian particles (ABPs)~\cite{Stenhammar2014} revealed the emergence of a steady-state \textit{bubble} phase, providing clear evidence of 
broken TRS with clusters stabilizing at a mesoscopic scale instead of spanning across the system. Interestingly, this behavior of ABPs contrasts with experimental observations in self-propelled colloids~\cite{Buttinoni2013, Theurkauff2012} where aggregation leads to the formation of \textit{clusters}.

The simplest extension of equilibrium Model B, which accounts for this broken TRS  is Active Model B (AMB), obtained by adding a gradient-nonlinear term to the equilibrium chemical potential,
$\mu=\delta\mathcal F/\delta\phi+\lambda(\nabla\phi)^2$~\cite{Wittkowski2014}. This term cannot be derived from a free-energy functional and therefore explicitly breaks detailed balance. While AMB captures key nonequilibrium features of motility-induced phase separation, renormalization-group(RG) analysis shows that it is not closed under coarse-graining, as additional active terms are generated, necessitating a more general model.
The minimal RG-consistent extension is Active Model B$+$ (AMB$+$), first emphasized by Tjhung \emph{et al.}~\cite{Tjhung2018}, with a dynamics given by
\begin{align}
    \label{eq: dphidt_AMB$+$}
    \partial_t \phi &= -\nabla\cdot\mathbf{J}+\sqrt{2DM}\mathbf{\Lambda} \nonumber\\
    \frac{\mathbf{J}}{M} &= -\nabla[\frac{\delta\mathcal{F}}{\delta\phi}+ \lambda|\nabla\phi|^2 + \frac{\nu}{2}\nabla^2(\phi^2)] \nonumber\\
    &~~~~+ \zeta\,(\nabla^2\phi)\nabla\phi,
\end{align}
where $\Lambda(\mathbf{r},t)$ is Gaussian white noise with zero mean and unit variance. $M$ is mobility and $D$ is thermal diffusion coefficient.
The negative gradient term in current $\mathbf{J}$ can be thought of as an effective  chemical potential $\mu = \meq +  \mu_{a}$ where the nonequilibrium chemical potential is generated by the active terms,  $\mu_{a} =\lambda|\nabla\phi|^2 + \frac{\nu}{2}\nabla^2(\phi^2)$. The equilibrium chemical potential $\meq=\delta\mathcal{F}/\delta\phi$ derived from a free energy functional  of  Model B, 
\begin{equation}
    \label{eq: Free_energy_functional}
    \mathcal{F}[\phi]=\int \left[f(\phi) + \frac\kappa2|\nabla\phi|^2\right] d\mathbf{r},
\end{equation}
with $f(\phi)=\frac r2\phi^2+\frac u4\phi^4$ being the bulk free energy density. The term $\zeta(\nabla^2\phi)\nabla\phi$ in $\mathbf{J}$ is the active current which can not be derived from a chemical potential, and it explicitly breaks TRS.

Together, all the active couplings $\lambda, \nu, \zeta$ break the $\phi \to -\phi$ invariance, and produce non-equilibrium currents.
Interestingly they all share the same canonical scaling dimension $(2-d)/2$, and are therefore marginal at $d=2$. A recent functional renormalization group (FRG) analysis by Fej\H{o}s, Sz\'ep, and Yamamoto~\cite{Fejos2025} clarifies the implications of this marginality. In two dimensions, the beta functions for $\lambda$ and $\nu$ lack linear terms and begin at quadratic order, so the renormalization-group flow does not approach a conventional infrared fixed point. Instead, the couplings evolve logarithmically slowly with scale: $\lambda(\ell),\nu(\ell), \zeta(\ell) \sim 1/\ln\ell$. Consequently, the effective mobility acquires scale dependence (i.e., $M\to M_{\text{eff}}(L)$), producing logarithmic corrections to correlation functions, as expected for systems governed by marginal operators at their upper critical dimension.
We propose that the implication of this marginality extends beyond criticality into the phase-separated regime and affects the coarsening dynamics substantially. A logarithmically slow evolution of couplings with scale may lead to effective interfacial mobility as $M_{\text{eff}}(L)=M_0(1 + \frac{b_0}{\ln(L/\xi)})$ which results in a modified growth law:$L(t) = L_0 t^{1/3}  (1 + \frac{c}{\ln t}).$

In this work, we revisit coarsening dynamics in AMB and AMB$+$ and investigate, through extensive numerical simulations, the possible emergence of logarithmic corrections to the growth law $L(t)$. We systematically analyze the interplay between different active terms and their collective influence on coarsening dynamics, demonstrating that the apparent deviation from the $t^{1/3}$ growth law can be consistently explained by the inclusion of suitable logarithmic corrections. Our results, therefore indicate that activity introduces marginal corrections to conserved diffusive coarsening rather than modifying the underlying scaling law itself.

The article is organized as follows. In Sec.~\ref{secII}, we study the critical behavior of AMB and AMB$+$ in the absence of any thermal noise and thereby verify mean-field predictions by extracting critical exponents through finite-size-scaling (FSS) analysis. 
Here we show that our order parameter faithfully reproduces the co-existence lines of phase separation transitions predicted theoretically. 
In Sec.~\ref{secIII}, we study the coarsening dynamics of both AMB$+$ and AMB, and find the existence of a slow logarithmic correction to the well-known $t^{1/3}$ power law, arising from the marginality of the activity parameters in $d=2$ to the well-known $t^{1/3}$ power law.
We conclude the results in Sec.~\ref{secIV}, with a short discussion on possible future directions.

\section{Critical Behavior}
\label{secII}
First we investigate the critical relaxation dynamics of both AMB and AMB$+$ in the absence of thermal noise $(D=0)$. 
Without noise,  and in the absence of  active terms, the dynamics is purely deterministic and governed by the mean-field equations of motion~\cite{Hohenberg1977}.  As a result,  the critical behavior in $d=2$ follow   the \textit{classical} mean-field exponents ~\cite{Kardar2007, Goldenfeld2018}. 
We ask how the active terms ($\lambda$, $\nu$, $\zeta$) modify the \textit{dynamical} critical exponent $z$ even at the mean-field level, or whether the critical relaxation remains identical to that of equilibrium Model B. 
By performing deterministic numerical simulations at the critical point $r_c=0$, we can isolate the effects of the non-linear gradient terms on the growth and saturation of the order parameter.

\subsection{Scaling and Critical exponents }
At the Gaussian (mean-field) fixed point, the static critical exponents take their classical values:
\begin{equation}
    \label{eq: MF_exponents}
    \quad \beta = \frac12,\quad \gamma = 1,\quad \eta = 0,\quad \nu = \frac12.
\end{equation}
The order parameter and the susceptibility scales as $\phi \sim |r|^{\beta}$ and  $\chi \sim |r|^{-\gamma}$ for $r<0.$ The critical exponent $\eta,\nu$ are associated with the equal-time correlation function and the correlation length, respectively. The active terms ($\lambda$, $\nu$, $\zeta$) alter the coarsening properties of the system by modifying the dynamics of the interfaces that separate the high and low density regimes~\cite{Nardini2017, Solon2018}, and one does not expect any change in the static critical exponents. Our aim would be to investigate if the dynamical critical exponents~\cite{Tuber2014} of AMB and AMB$+$ are modified.
\begin{figure}[t]
    \centering    \includegraphics[width=\linewidth]{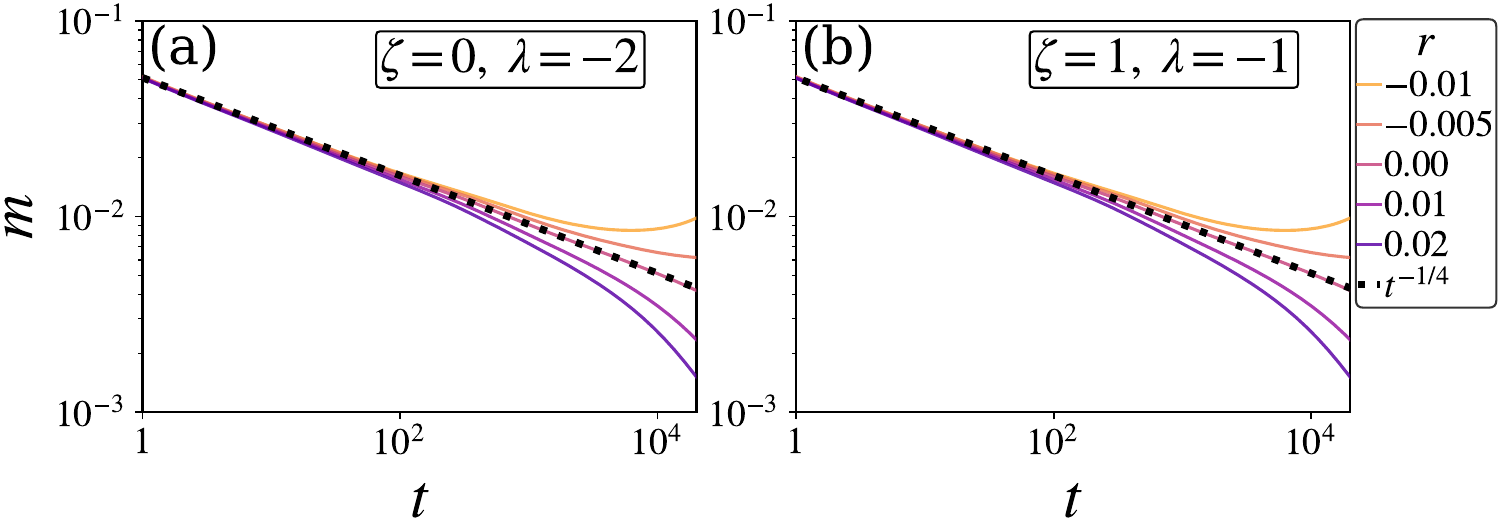}
    \caption{Decay of order parameter $m(t)$  for (a) AMB: $\zeta=0,\lambda=-2,\nu=0.5$, and (b) AMB$+$: $\zeta=1,\lambda=-1,\nu=0.5$. In super critical regime ($r<0$), $m$ saturates to a steady value $m_{\rm s}$ whereas it vanishes in the subcritical ($r>0$) region. At the critical threshold i.e. $r_c=0$, $m(t) \sim  t^{-\alpha}$, both AMB and AMB$+$ exhibit critical decay exponent $\alpha=\frac14$ (which is the slope of the dashed line). Here, the system size considered is $L=256$.}
    \label{fig: beta/nu_t extraction}
\end{figure}
Critical behaviour and finite-size scaling follow standard scaling theory
\cite{Stanley1971,Cardy1996,Privman1990}.
In equilibrium Model B ($\lambda=0=\nu, \zeta=0$), the purely relaxational deterministic equation at the critical point ~\cite{Bray1994}
\begin{equation}
    \label{eq: relaxation_eqn}
    \partial_t \phi = -\kappa \nabla^4 \phi + \mathcal{O}(\phi^3, \nabla^2\phi^3, \dots),
\end{equation}
gives the scaling relation $[t] = [x]^4$, resulting in a growth of correlation length $\xi$ as $t^{1/z}$ with the dynamic exponent is $z = 4$. Another critical exponent associated with the dynamics is $\nu_t$, which  governs the scaling of the relaxation time $\xi_t$ near criticality as $\xi_t \sim |r|^{-\nu_t}$, and obeys a scaling relation $\nu_t=z\nu$ \cite{Stanley1971, Cardy1996}. Thus from Eq.~\eqref{eq: MF_exponents} we have $\nu_t=\frac12$.

To study the critical behavior in AMB/AMB$+$, we need to define a suitable order parameter. Since, the scalar field $\phi(\mathbf r,t)$   can be interpreted as the deviation of the coarse-grained local density $\rho({\bf r},t)$ from the conserved global density $\bar \rho$ of the system~\cite{Chaikin1995, Tjhung2018}
\begin{equation}
    \label{eq: phi_rho_transformation}
    \phi(\mathbf r,t) = \rho(\mathbf r,t)-\bar\rho,
\end{equation}
clearly an inhomogeneous density profile results in $\phi\ne0$: 
the high and low-density regions are represented by $\phi>0$ and $\phi<0$ respectively.
Thus, one can define the order parameter of the system to be
\begin{equation}
    \label{eq: order_parameter}
    m(t,L)  \equiv  \frac{1}{L^d} \int d^d r \, |\phi(\mathbf{r},t)|,
\end{equation}
which vanishes in the mixed phase and picks up non-zero values in the phase-separated state. A similar order parameter is defined for MIPS transition in lattice models  of run and tumble particles \cite{Ray2024}.

In a finite system of linear size $L$, the critical fluctuations at $r_c=0$ are cut off by the finite system size, leading to finite-size scaling (FSS)~\cite{Stanley1971, LandauBinder2014},      
\begin{equation}
    \label{eq: FSS}
m(t,L) = t^{-\alpha} \, f\bigl(t / L^{z}\bigr),
\end{equation}
$z$ being the dynamical exponent at criticality.
Since the correlation length $\xi$ and time $\xi_t$ scales at criticality as $\xi \sim \varepsilon^{-\nu}$ and $\xi_t \sim \varepsilon^{-\nu_t}$ respectively with $\varepsilon=r_c-r$, one expects  $\xi_t \sim \xi^{\nu_t/\nu}$. Thus, for a finite system, we have $t \sim L ^{\nu_t/\nu}$, resulting in $z=\nu_t/\nu$. Also, in the supercritical steady state (i.e. in $t\to \infty$ limit) $m$ must saturate to a value $m_s$, proportional to $|\epsilon|^\beta \equiv  L^{\beta/\nu}$. Thus, 
\begin{equation}
    \label{eq: m_L_scaling}
    m(t,L) = \begin{cases}
                t^{-\alpha} & \text{as } L \to \infty \\
                L^{-\beta/\nu} & \text{as } t \to \infty.
            \end{cases}
\end{equation}
Now, for the scaling function $f(\cdot)$ in Eq.~\eqref{eq: FSS} to be consistent with above, it must obey, 
\begin{align}
    \label{eq: Scaling_Function}
    f(x) =
        \begin{cases}
            x^{-\beta/(z\nu)} & \text{for } x \to 0 \\
            \text{constant} & \text{for } x \to \infty.
        \end{cases}
\end{align}
Thus, $\alpha$ must be same as $\frac{\beta}{z\nu}=\frac{\beta}{\nu_t}.$

We first analyze the time evolution of the order parameter $m(t,L)$ to extract the critical exponent $\alpha$. The numerical procedure  for integrating Eq.~\eqref{eq: dphidt_AMB$+$} is described in Appendix \ref{apndx: A}. For a system size of $L=256$, we plot $m(t,L)$ versus $t$ (in log-log scale) at different $r$ values, \textit{viz}. at $r=-0.01,-0.005,0,0.01,0.02$ in Fig.~\ref{fig: beta/nu_t extraction}. For $r>0$ the curves decay faster than power laws, while for $r<0$ they saturate, indicating a critical point at $r_c=0$. Using the critical scaling relation $m(t,L)\sim t^{-\alpha}$ at $r_c$, a linear fit in log-scale yields $\alpha=0.25$. The exponent values obtained for both AMB $(\zeta=0,\lambda=-2,\nu=0.5)$ in Fig.~\ref{fig: beta/nu_t extraction}(a) and for AMB$+$ $(\zeta=1,\lambda=-1,\nu=0.5)$ in Fig.~\ref{fig: beta/nu_t extraction}(b), are same.

We must mention that the critical value $r_c$  also suffers from the finite-size effects; the finite-size critical value $r_c(L)$ is slightly smaller than its thermodynamic limit $r_c(\infty)=0$. This shift follows directly from the linear dispersion relation
\begin{equation}
    \label{eq: omega}
    \omega(k) = -k^2\left(r + \kappa k^2\right),
\end{equation} which implies that the homogeneous density profile is linearly stable only when $r + \kappa k^2 > 0$. In a finite system with periodic boundary conditions, the wave vectors are discrete and the smallest nonzero mode is $k_{\min}=2\pi/L$. Stability must therefore hold when $r + \kappa k_{\min}^2 > 0$, yielding

\begin{equation}
    \label{eq: r_c_finite_size}
    \quad r_c(L) = -\kappa \left(\frac{2\pi}{L}\right)^2.
\end{equation}
Thus, the shift of the critical point follows the finite-size scaling form $r_c(L)-r_c(\infty)\sim L^{-1/\nu}$, with $\nu=2$.

Equation~\eqref{eq: r_c_finite_size} shows that a finite system at  $r=0$ is effectively subcritical, since the true critical point is shifted to $r_c(L)<0$. Consequently, the order parameter $m(t,L)$ ultimately decays faster than a power law at long times. 
An extended algebraic regime $m(t)\sim t^{-\alpha}$ is therefore observed only for sufficiently large $L$ (like $L=256$ in Fig.~\ref{fig: beta/nu_t extraction}) or when dynamics are analyzed relative to $r_c(L)$.
\begin{figure}[t]
    \centering
    \includegraphics[width=\linewidth]{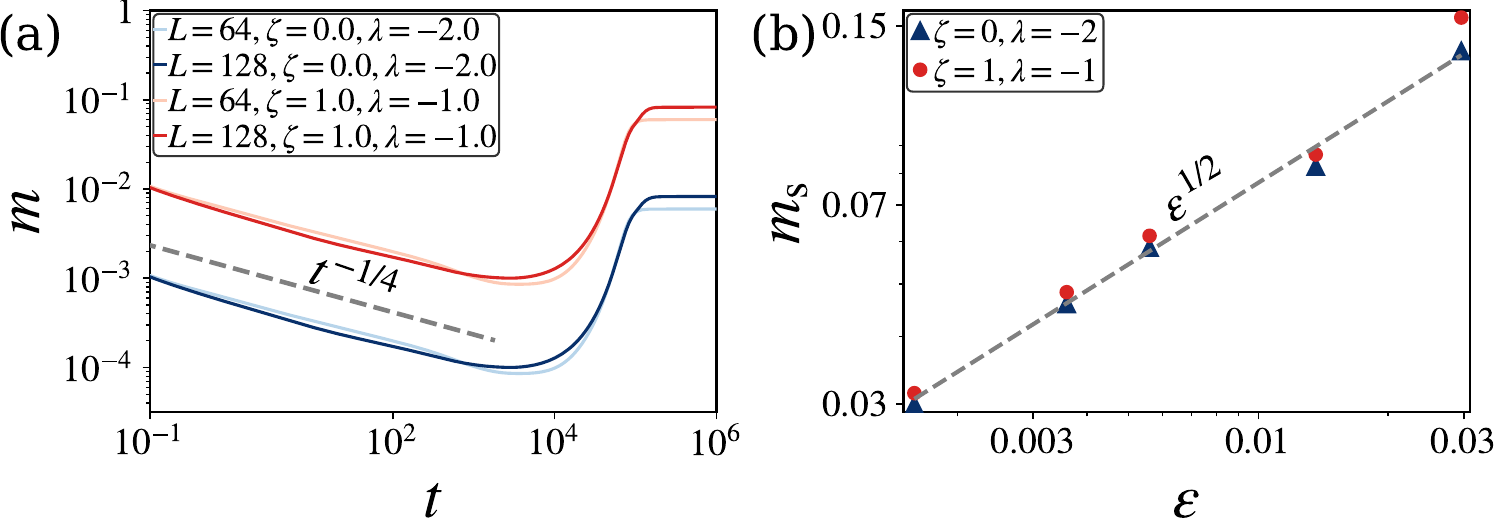}
    \caption{(a) Log-log plot of order parameter $m$ with time $t$ in the super-critical region ($r<0$), for  AMB ($\zeta=0,\lam=-2,\nu=0.5$--lower curves) and AMB$+$ ($\zeta=1,\lam=-1,\nu=0.5$--upper curves). We consider two system sizes: $L=64$, and $128$ at $r=-0.016$, which is well below their respective $r_c(L)$ given in Eq.~\eqref{eq: r_c_finite_size}. (b) Log-log plot of $m_s$ vs $\epsilon= r_c(L)-r$: for both AMB and AMB$+$,  $m_s\sim \epsilon^\beta$ gives $\beta=\frac12$. Here, $L=128$ and data for AMB$+$ is multiplied by a factor for better visual clarity.}
    \label{fig: beta}
\end{figure}

To determine the order parameter exponent $\beta$, we analyze the supercritical regime $r<r_c$, where the order parameter $m(t,L)$ saturates at long times to a finite value
\begin{equation}
    \label{eq: m_vs_eps}
    m_{\rm s}\sim \varepsilon^\beta, 
    \qquad \varepsilon=(r_c-r).
\end{equation}
We perform simulations at system size $L=128$, for which $r_c=-(2\pi/128)^2$ (with $\kappa=1$), and measure the steady-state value $m_{\rm s}$ for several values of $r$. In Fig.~\ref{fig: beta}(a) we show the saturation of $m(t,L)$ with $t$ in log-log plot for both AMB ($\zeta=0,\lambda=-2,\nu=0.5$) and AMB$+$($\zeta=1,\lambda=-1,\nu=0.5$) at an arbitrary value of $r=-0.016$, which is well below the respective $r_c(L)$ for $L=64$ and $128$. For both cases, the curves initially decay as $t^{-1/4}$, but before saturation they exhibit undershooting. A log-scale plot of this $m_{\rm s}$ as a function of $\varepsilon=(r_c-r)$, with $r_c>r$ is shown in Fig.~\ref{fig: beta}(b) for both AMB and AMB$+$. In both cases we find that $\beta=\frac12$ fits the data quite well.

\begin{figure}[t]
    \centering
    \includegraphics[width=\linewidth]{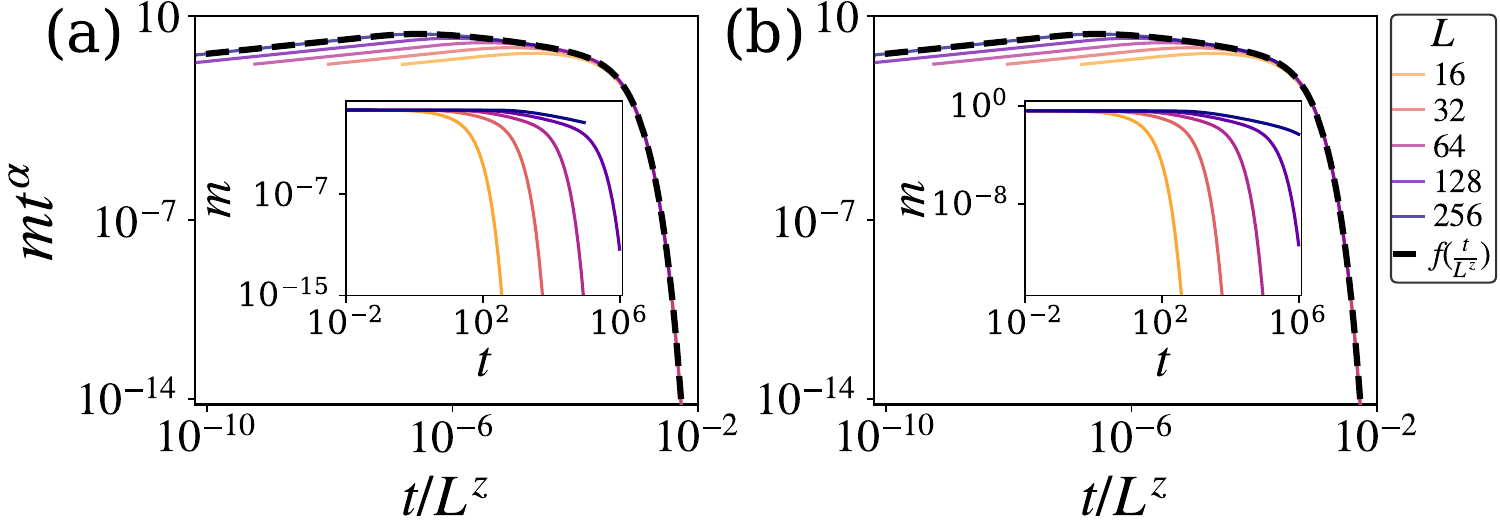}
    \caption{(a) Scaling collapse in AMB ($\zeta=0,\lambda=-2, \nu=0.5, r=0$): log-log plot of $m(t)t^{-\alpha}$ as a function of $t/L^z$ for different $L$ collapse onto each other when $\alpha=\frac14, z=4$. (b) The same for AMB$+$ ($\zeta=1,\lam=-1,\nu=0.5$).  For both AMB and AMB$+$, the collapsed curve matches the corresponding scaling function of equilibrium Model B (dashed line) after suitable re-scaling  of the $x$-axis.}
    \label{fig: m vs. t scaling collapse}
\end{figure}

Using the critical exponents $\beta=\nu=\frac12$ and $\nu_t=2$, we test the scaling relation in Eq.~\eqref{eq: FSS}. We compute $m(t,L)$ for different system sizes and plot $m t^{\alpha}$ versus $t/L^z$ with $\alpha=\tfrac14$ and $z=4$. Excellent data collapse is observed for both AMB $(\zeta=0,\lambda=-2,\nu=0.5)$ (Fig.~\ref{fig: m vs. t scaling collapse}(a)) and AMB$+$ $(\zeta=1,\lambda=-1,\nu=0.5)$ (Fig.~\ref{fig: m vs. t scaling collapse}(b)). The dashed lines in Fig.~\ref{fig: m vs. t scaling collapse}(a) and ~\ref{fig: m vs. t scaling collapse}(b) is the scaling function $f(.)$ obtained for the equilibrium Model B, which matches well with that of the AMB and AMB$+$. This confirms that (i) $m(t)$, defined in Eq.~\eqref{eq: order_parameter}, serves as an appropriate order parameter for the phase-separation transition, and (ii) the transition in AMB and AMB$^+$ belongs to the same universality class as equilibrium Model B. This finding is consistent with lattice models~\cite{Partridge2019, Ray2024} of active particles, which exhibit Ising universality in two dimensions.

\subsection{Phase diagram: Coexistence line}

In this section, we determine the binodal densities $\phi_-$ (gas/vapor phase) and $\phi_+$ (liquid/dense phase), defined as the coexisting steady state densities across a flat interface. With $\phi>0$ representing the high-density region and $\phi<0$ the low-density region, the corresponding areas can be computed numerically as
\begin{equation}
    \label{eq: Area_fraction}
    A_{\pm} = \frac{1}{L^d} \int d^d r \, \Theta\left( \pm \phi(\mathbf{r},t) \right) 
\end{equation}
where $\Theta(.)$ is the Heaviside step function. Now, the binodal densities $\phi_\pm$, can be defined as 
\begin{equation}
    \label{eq: phi_pm}
    \phi_{\pm} = \frac{\Phi_\pm}{A_{\pm}}; \, 
   \Phi_\pm= \int d^d r \,
     \phi(\mathbf{r},t)\Theta\left( \pm\phi(\mathbf{r},t) \right).
\end{equation}

To calculate $\phi_\pm$ analytically, we follow the approach prescribed in Refs.~\cite{Wittkowski2014, Solon2018, Solon2018IOP}. For the noiseless ($D=0$) case of Eq. \eqref{eq: dphidt_AMB$+$}, if we consider a flat interface, then clearly the density profile depends only on the normal coordinate $x$ (say). Now, in this one-dimensional ($d=1$) system, using identities: 
$i)\;\frac{\nu}{2}\partial_{xx}\phi^2=\nu(\partial_x\phi)^2+\nu\phi\partial_{xx}\phi$ and 
$ii)\;\partial_x(\partial_x\phi)^2=2\partial_x\phi\,\partial_{xx}\phi$, 
the coefficients of the terms $\partial_{xx}\phi$ and $(\partial_x\phi)^2$ in Eq.~\eqref{eq: dphidt_AMB$+$}, become $K(\phi)=\kappa-\nu\phi$ and $\tilde{\lambda}=\lambda+\nu-\zeta/2$ respectively. The problem thus reduces to a one-dimensional gradient current $J=-\partial_x\mu$, with (setting  $M=1$) an effective nonequilibrium chemical potential
\begin{equation}
    \label{eq: Binodal_chem_pot}
    \mu=f'(\phi)-K(\phi)\partial_{xx}\phi+\tilde{\lambda}(\partial_x\phi)^2 ,
\end{equation}
where $f'(\phi)=r\phi+u\phi^3$. This chemical potential has the same form as in Ref.~\cite{Tjhung2018}, with $\tilde{\lambda}$ replacing $\lambda-\zeta/2$.

We seek a stationary profile $\phi(x)$ (with $u=1$) satisfying $\phi(-\infty)=\phi_-$ and $\phi(+\infty)=\phi_+$, describing a smooth interface centered at $x=0$. In the bulk, homogeneity implies $J=0$, so the chemical potential should remain constant there, say $\mu=\mu_0$, yielding the first coexistence condition
\begin{equation}
    \label{eq: 1st_relation}
    f'(\phi_-)=f'(\phi_+)=\mu_0 .
\end{equation}
This relation alone does not determine the binodals. Following \cite{Solon2018,Solon2018IOP}, we introduce a pseudodensity $R(\phi)$ and consider the integral $I=\int_{x_1}^{x_2}\mu\,\partial_x R(\phi)\,dx$. Using $\mu(x)=\mu_0$, the integral takes the implicit form
\begin{equation}
    \label{eq: Interface_Integral_implicit}
    I = \mu_0\big(R(\phi_+) - R(\phi_-)\big).
\end{equation}
Due to the presence of the gradient terms in $\mu$ as in Eq.~\eqref{eq: Binodal_chem_pot}, computing $I$ explicitly is generally difficult. For the specific form of the pseudodensity $R(\phi)$, such that the relation
\begin{equation}
    \label{eq: R_equation}
    K R^{''} = -(2\tilde{\lambda} + K^{'})R^{'}
\end{equation}
holds, where ${'}\equiv\partial_{\phi}$, this can be done. Then the integral explicitly gives
\begin{equation}
    \label{eq: Interface_Integral_explicit}
    I = g(\phi_+) - g(\phi_-),
\end{equation}
where the pseudopotential $g(\phi)$ satisfies the relation $\frac{\partial g}{\partial R}=\frac{\partial f}{\partial\phi}$.

Then, Eqs.~\eqref{eq: Interface_Integral_implicit} and \eqref{eq: Interface_Integral_explicit} enforces the equality of the pseudopressure, defined as $P=R\frac{\partial g(\phi)}{\partial R}-g(\phi)$ at the interface between the two phases, giving a second relation,
\begin{equation}
    \label{eq: 2nd_relation}
    R(\phi_-)\mu_0(\phi_-) - g(\phi_-) = R(\phi_+)\mu_0(\phi_+) - g(\phi_+).
\end{equation}
With the specific form of the chemical potential as in Eq.~\eqref{eq: Binodal_chem_pot}, the explicit form of $R$ and $g$ is calculated for the case of AMB$+$, to be
\begin{align}
    \label{eq: AMB_R_f}
    R & = -\frac{1}{\nu}\frac{y^{1+a}}{(1+a)}\\
    g & = \frac{1}{\nu^4(4 + a)}y^{4+a} - \frac{3\kappa}{\nu^4(3 + a)}y^{3+a} \notag\\
    & +\frac{3\kappa^2/\nu^4 + r/\nu^2}{(2+a)}y^{2+a} - \frac{\kappa^3/\nu^4 + r\kappa/\nu^2}{(1+a)}y^{1+a}
\end{align}
where $y=\kappa-\nu\phi$, and $a=\frac b\nu$ with $b=2\lambda + \nu - \zeta$.

The binodal densities are determined as stationary solutions ($\phi_{\pm}$) by enforcing equality of $g_0$ and $P$ across the two coexisting phases. Numerically, it can be obtained by simultaneously solving Eqs.~\eqref{eq: 1st_relation} and \eqref{eq: 2nd_relation}, which is  equivalent to a common-tangent construction on $\phi(R)$.

\begin{figure}[t]
    \centering
    \includegraphics[width=\linewidth]{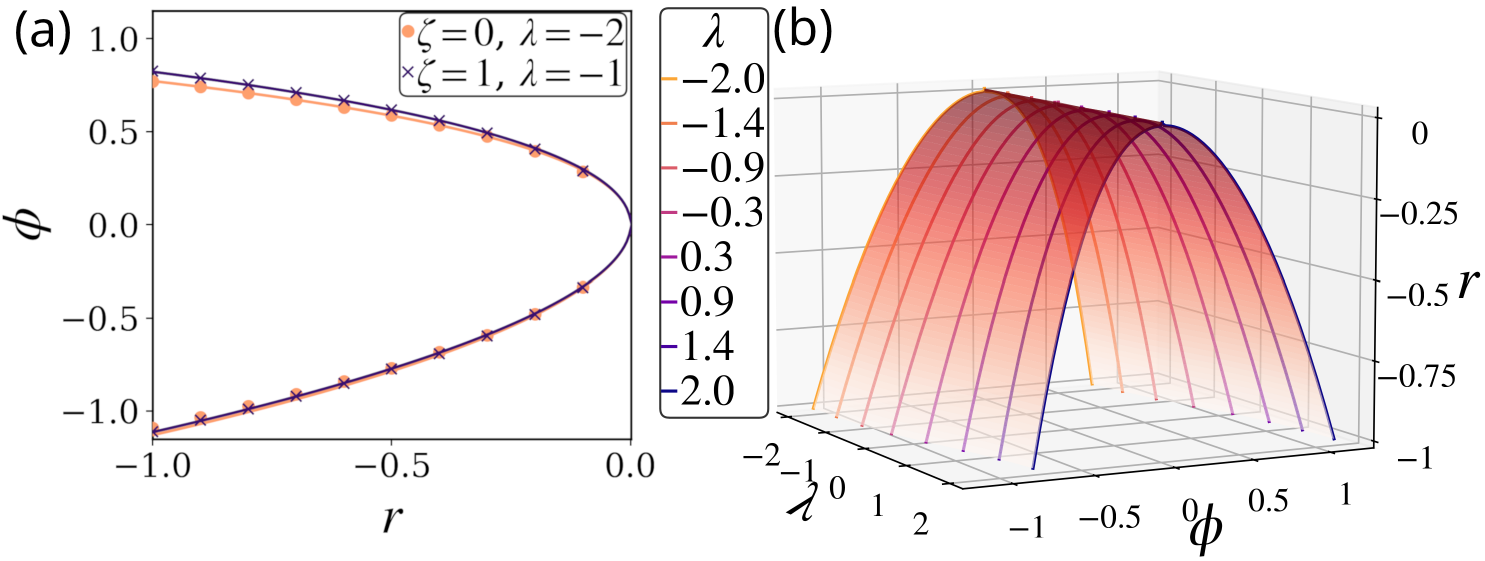}
    \caption{(a) Binodal diagram in the $r-\phi$ plane for AMB ($\zeta=0, \lambda =-2, \nu=0.5$) and AMB$+$($\zeta=1, \lambda =-1, \nu=0.5$).The solid lines show the exact analytical values of $\phi_+$ (upper branch) and $\phi_-$ (lower branch), obtained from Eqs.~\eqref{eq: 1st_relation} and \eqref{eq: 2nd_relation}.  Corresponding points are obtained from simulating the noiseless PDE numerically on a $128\times128$ grid and using  Eqs.~\eqref{eq: Area_fraction} and  \eqref{eq: phi_pm}. (b) Full three dimensional phase-diagram of phase separation transition in AMB$+$ in $\phi$-$\lambda$-$r$ plane for $\zeta=1,\nu=0.5$.}
    \label{fig: phase_diagram}
\end{figure}

To obtain $\phi_\pm$ numerically, we use Eqs. \eqref{eq: Area_fraction} and ~\eqref{eq: phi_pm}. In steady state, we compute the area fractions $A_\pm$ as the number of grid points with $\phi(\mathbf r,t)\gtrless0$, and $\Phi_\pm$ as the corresponding sums of $\phi(\mathbf r,t)$. Then the coexistence densities can be obtained as $\phi_\pm=\Phi_\pm/A_\pm$. A binodal diagram in $r$--$\phi$ plane is shown in Fig.~\ref{fig: phase_diagram}(a) for AMB $(\zeta=0,\lambda=-2,\nu=0.5)$ and AMB$+$ $(\zeta=1,\lambda=-2,\nu=0.5)$, showing excellent agreement with analytical predictions, therefore validating the order parameter of Eq.~\eqref{eq: phi_pm}. Figure~\ref{fig: phase_diagram}(b) displays the binodal surface in the extended $\phi$--$\lambda$--$r$ space at fixed $\zeta=1$. Each curve represents $\phi_\pm(r)$ for a given $\lambda$, forming a smooth phase-separated surface. Varying $\lambda$ mainly changes the magnitude of $\phi_\pm$, i.e., the binodal width, rather than shifting the critical point. The apex near $r=0$ remains nearly unchanged, while symmetry about $\phi=0$ is broken, reflecting the breaking  of time-reversal symmetry in AMB and AMB$+$ models.

\section{Super-critical  quench}
\label{secIII}

To understand the coarsening process in AMB$+$, first we look at how the dynamics in Eq.~\eqref{eq: dphidt_AMB$+$} scales under naive renormalization group approach: $x\to bx,~ t\to b^{z}t$ and $\phi\to b^{\chi}\phi$. The dominant linear term near criticality remains $\partial_t\phi \sim -\kappa\nabla^4\phi$ and matching their scaling dimensions gives the Gaussian dynamical exponent $z=4$. The field scaling follows from the quadratic free-energy functinal part $\mathcal F\sim\int d^d x\left(\nabla\phi\right)^2$, yields
$\chi=\frac{2-d}{2}$. The active nonlinearities $\nabla^2(\nabla\phi)^2$, $\nabla^2(\phi\nabla^2\phi)$ scale as $\sim b^{2\chi-4}$ and 
comparing it with the  scaling of linear-terms $b^{\chi-4}$ shows that the scaling dimension of the couplings $\lambda$, $\zeta$, and $\nu$ are  $[\lambda]=[\zeta]=[\nu]=-\chi=\frac{d-2}{2}$. Hence,  at the Gaussian fixed point, they are marginal in $d=2$, irrelevant for $d>2$, and relevant for $d<2$ . Thus naive power counting for AMB$+$ leads to the fact that the activity enters as a marginal perturbations in two dimensions; only higher-order fluctuations can determine whether these operators are marginally relevant or irrelevant.

Marginal perturbations typically do not modify universal power-law exponents but generate multiplicative logarithmic corrections to scaling. Classic examples include the $\phi^4$ interaction at the upper critical dimension $d_c=4$~\cite{Wilson1974,Cardy1996}, logarithmic corrections in the two-dimensional XY model near the Kosterlitz--Thouless transition~\cite{Kosterlitz1973}, and conserved phase-ordering kinetics at the marginal dimensions~\cite{Bray2002}. Such logarithmic factors reflect slow RG flow near the fixed point rather than the emergence of new scaling exponents.

To understand coarsening in AMB$+$, we first recall the passive case of equilibrium Model B, with a $\phi^4$ free energy. Coarsening is driven by interfacial energy reduction, where the equilibrium chemical potential $\meq=\delta\mathcal F/\delta\phi$ is dominated by the Laplace pressure contribution, with scaling with the curvature as $\kappa\sim1/L$~\cite{Bray2002,Cates2017}. For conserved dynamics $\partial_t\phi=\nabla^2\meq= -\nabla \cdot \mathbf J$, where the current $\mathbf J=-M\nabla\meq$ scales as $\mathbf J\sim M/L^2$, with $M$ being the mobility. The change of domain size is now driven by the current, leading to $dL/dt\sim M/L^2$, or equivalently $L(t)\sim t^{1/3},$ the Lifshitz--Slyozov growth law.

For AMB and AMB$+$ in two dimension, we expect the growth law to be determined by a balance:
\begin{equation}
     \label{eq: scaling_balance}
     \frac{dL}{dt} \sim \frac{M_{\text{eff}}(L)}{L^2},
\end{equation}
where $M_{\text{eff}}(L)$ is an effective, scale-dependent mobility, in contrast to the constant $M$ of equilibrium  Model B. The active terms, being marginal in nature depends weakly (more precisely logarithmically) on the domain size $L$
\begin{equation}
    \label{eq: M_eff}
    M_{\text{eff}}(L) = M_0 \left[ 1 +  \frac{b_0}{\ln(L/\xi)} + \dots \right],
\end{equation}
where $\xi$  is the correlation length  providing a  microscopic cutoff. Thus, to the leading order in $L$ we can write
\begin{equation}
    \label{eq: growth_eq}
    \frac{dL}{dt} \sim \frac{1}{L^2} \left[ 1 + \frac{3 c}{\ln(L)} \right],
\end{equation}
where the constant $3c$ absorbs the parameters $b$ and $\xi$. This can be solved asymptotically, by writing $L(t) =  t^{1/3} \Psi(t)$, where $\Psi(t)$ is a slowly varying  function of $t.$  Expanding at long time $t$, one finds that to the leading order in $\ln(t)$:
\begin{equation}
    \label{eq: nonLS}
    L(t)=L_0 t^{1/3}(1 + \frac{c}{\ln t}).
\end{equation}
Clearly, for very large $t$ one recovers the standard Lifshitz--Slyozov (LS) growth law $L(t)\sim t^{1/3}$.

The coarsening behavior of AMB and its extension AMB$+$ has been investigated previously by several authors~\cite{Stenhammar2013, Wittkowski2014,Tjhung2018,Pattanayak2021,Yadav2025}. These works established AMB/AMB$+$ as continuum field theories for scalar active matter (see Eq.~\eqref{eq: dphidt_AMB$+$}) that break time-reversal symmetry while conserving density. Numerical simulations revealed domain growth broadly consistent with diffusion-limited coarsening, but accompanied by unusually slow relaxation and strong corrections to scaling, which obscured a clear identification of the asymptotic growth law~\cite{Stenhammar2013, Wittkowski2014}.

Subsequent large-scale numerical studies~\cite{Pattanayak2021, Yadav2025} reported that the characteristic domain size appeared to follow an effective power law $L(t)\sim t^{1/z_c},$ 
with a dynamic exponent $z_c$ varying continuously with activity parameters. This behavior suggested a possible violation of classical LS scaling for conserved coarsening. Here,  since the critical behavior of AMB and AMB$+$ coincides with that of equilibrium Model B, rather than invoking a new scaling law, we propose that the observed drift of effective exponents originates from strong logarithmic corrections to LS scaling. We demonstrate that the numerical data are well described by Eq.~\eqref{eq: nonLS}, which preserves the LS asymptotic exponent while naturally producing extended pre-asymptotic regimes that mimic continuously varying effective exponents.

\subsection{Coarsenning in Active Model B ($\zeta = 0$)}
\label{subsec: AMB_log}

\begin{figure}[t]
    \centering
    \includegraphics[width=0.5\textwidth]{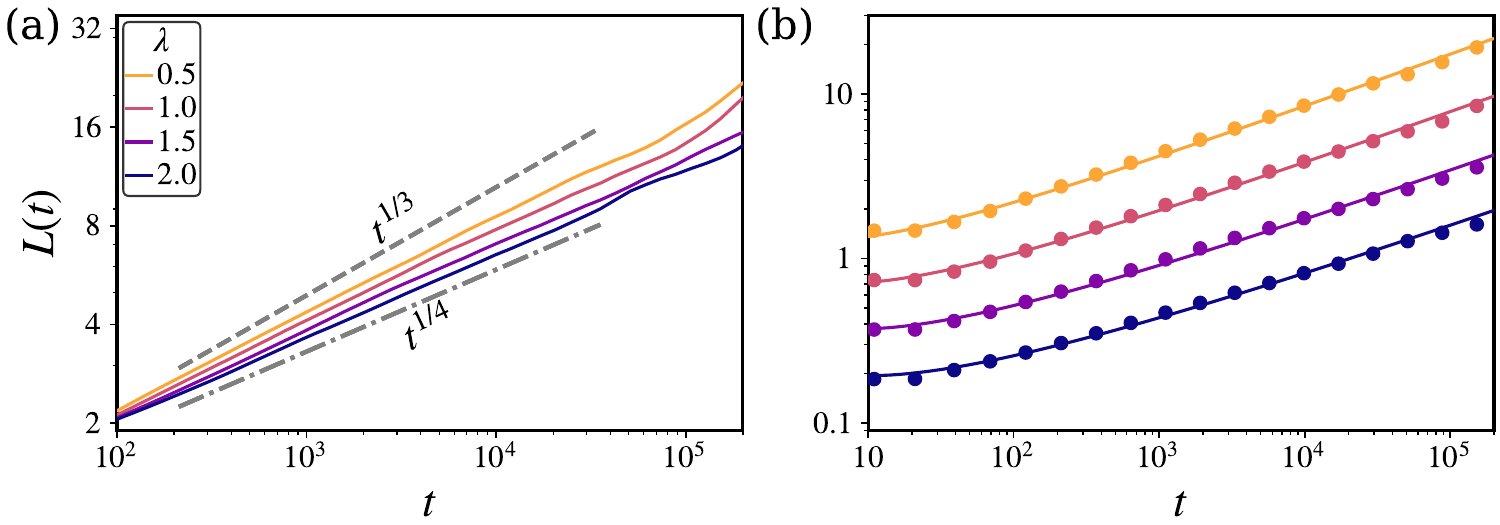}
    \caption{Active Model B ($\zet = 0$) shows logarithmic corrections. (a) The coarsening length $L(t)$ as a function of time for various $\lam$ values. While the raw data appear to follow a power law $t^{\theta}$, with $\theta$ varying continuously with $\lambda$ and approaching $\frac13$ as $\lambda \to 0.$ (b) The same data (symbols) fits to theoretical predictions (lines) as in Eq.~\eqref{eq: nonLS} $c=2.325, 3.661, 5.688, 7.75$ for $\lambda=0.5,1,1.5,2$ respectively.
    The curves are vertically shifted for  clarity. Here $\zeta=0, \nu=0.5, r=-1,$ and  the system size is $L=256.$}
    \label{fig: AMB_log_correction}
\end{figure}
Numerical evidence for the presence of logarithmic correction in the coarsening length $L(t)$ in the case of AMB is presented in Fig.~\ref{fig: AMB_log_correction}.
To measure $L(t)$, we compute the equal-time two-point correlation function
$C(\mathbf{r}, t) = \left\langle \phi(\mathbf{R}, t)\,\phi(\mathbf{R} + \mathbf{r}, t) \right\rangle 
- \left\langle \phi(\mathbf{R}, t) \right\rangle 
\left\langle \phi(\mathbf{R} + \mathbf{r}, t) \right\rangle$. Then, the corresponding equal-time structure factor
$S(\mathbf{k},t)$ is computed by taking the Fourier transform of $C(\mathbf{r},t)$. The characteristic domain size $L(t)$ is then defined as
\begin{equation}
    \label{eq: L(t) from S(k)}
    L(t) =\frac{\int  S(k,t)dk}{\int k S(k,t)dk},
\end{equation}
where $k=\lVert \mathbf{k} \rVert$ is the modulus of wave vector $\mathbf k$ and $S(k,t)=\langle \phi(\mathbf{k},t)\phi(-\mathbf{k},t) \rangle_k$ is the spherically averaged structure factor.

The measured values of $L(t)$ for different $\lam$ are shown as a function of $t$ in Fig.~\ref{fig: AMB_log_correction}(a) in log scale. For small $t$, it appears that $L(t) \sim t^{\theta}$ with the exponent $\theta$ varying continuously with $\lambda$ and systematically $\theta$ approaches $1/3$ as $\lambda\to 0$, similar to earlier observations as in Ref.~\cite{Wittkowski2014}. In Fig.~\ref{fig: AMB_log_correction}(b), we plot $L(t)$ as a function of $t$ for different $\lam$ values (symbols), along with the logarithmic scaling form of $L(t)$ given in Eq.~\eqref{eq: nonLS}. Considering $c$ as a fitting parameter, we find a best-fit to the data at  $c=2.325, 3.661, 5.688, 7.75$ respectively for $\lam=0.5, 1, 1.5, 2$ respectively. It can be seen from Fig.~\ref{fig: AMB_log_correction}(b), that the data fits very well to Eq.~\eqref{eq: nonLS}, indicating an alternative scenario consistent with theoretical prediction. This confirms that  the $\lam$ term is marginally relevant and  its primary effect is to introduce a slow, logarithmic correction to the classic $t^{1/3}$ law.

\subsection{Coarsenning in Active Model B$+$}
\label{subsec: AMBplus_crossover}

We study the coarsening dynamics of Active Model B$+$ (AMB$+$), where the additional activity parameter $\zeta$ generates higher–order, non-integrable interfacial currents that break time-reversal symmetry beyond AMB~\cite{Wittkowski2014}. These active currents compete with diffusive transport responsible for classical \textit{Cahn}--\textit{Hilliard} coarsening and can qualitatively modify phase-separation kinetics, potentially reversing Ostwald ripening and stabilizing microphase-separated states with a finite characteristic length scale. In this work, we focus on the positive $\zeta$ regime, fixing $\zeta=1$, where coarsening persists over long times while retaining clear nonequilibrium signatures. This parameter range allows us to probe how activity modifies domain growth laws and to test deviations from classical Lifshitz--Slyozov (LS) scaling without entering the fully arrested bubbly phase as mentioned in Ref.~\cite{Tjhung2018}.

\begin{figure}[t]
    \centering
    \includegraphics[width=0.5\textwidth]{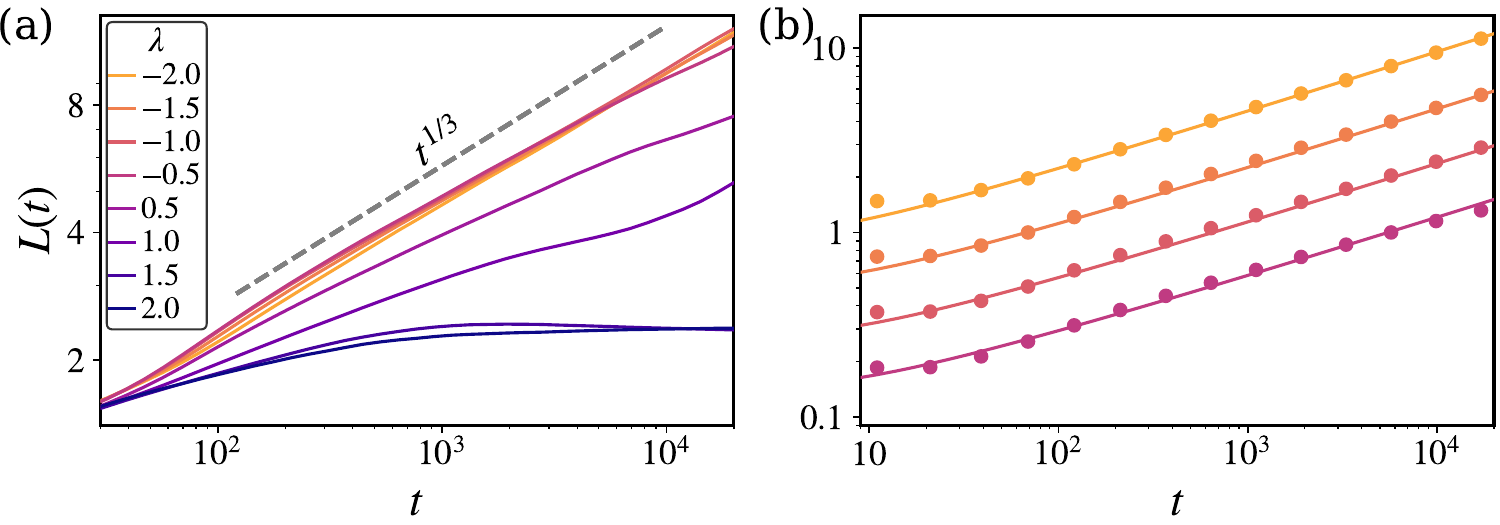}
    \caption{Active Model B$^+$ ($\zeta=1$) exhibits weak logarithmic corrections to coarsening. (a) Log--log plot of $L(t)$ versus $t$ for different values of $\lambda$. For $\lambda<0$, the growth remains close to the Lifshitz--Slyozov law $L(t)\sim t^{1/3}$, with small deviations attributable to logarithmic correction. For large positive $\lambda$, $L(t)$ rapidly saturates, indicating a microphase-separated state associated with reverse Ostwald ripening. Intermediate $\lambda$ values show anomalous growth, requiring larger system sizes for reliable characterization. (b) Data for $\lambda<0$ (symbols) are well described by Eq.~\eqref{eq: nonLS} with $c=1.3,\,1.2,\,1.1,$ and $0.8$ for $\lambda=-0.5,-1,-1.5,$ and $-2$, respectively. Parameters are $\zeta=1$, $\nu=0.5$, $r=-1$, system size $L=1024$, and curves are vertically shifted for clarity.}
    \label{fig: AMB$+$_log_correction}
\end{figure}

To characterize the coarsening behavior, we numerically integrate the AMB$+$ equation i.e. Eq.~\eqref{eq: dphidt_AMB$+$}, on a $L\times L$ square-grid with $L=1024$, keeping $\nu=0.5$ and $\zeta=1$ fixed while varying $\lambda$ in the range $-2\le\lambda\le2$. The characteristic domain size $L(t)$ is extracted from the structure factor according to Eq.~\eqref{eq: L(t) from S(k)} and shown on logarithmic scales in Fig.~\ref{fig: AMB$+$_log_correction}.

We first consider the case $\lambda<0$. As shown in Fig.~\ref{fig: AMB$+$_log_correction}(a), for $t \gtrsim 10^2$ the domain size $L(t)$ follows the classical $t^{1/3}$ growth law with only weak deviations. These small departures are consistent with the logarithmic correction proposed in Eq.~\eqref{eq: nonLS}. Indeed, Fig.~\ref{fig: AMB$+$_log_correction}(b) demonstrates that the data are well described by the form of Eq.~\eqref{eq: nonLS} over the entire temporal range, including relatively early times. The fitted coefficients $c=1.3,\,1.2,\,1.1,$ and $0.8$ for $\lambda=-0.5,-1,-1.5,$ and $-2$, respectively, are significantly smaller than those obtained for AMB, indicating weaker corrections to LS scaling. This suggests that the presence of the $\zeta$ term stabilizes curvature-driven coarsening by reducing the slow transient effects induced by the $\lambda$ term. We therefore hypothesize that for $\lambda<0$ the combined action of the $\lambda$ and $\zeta$ terms suppresses anomalous interfacial currents, restoring nearly classical diffusive coarsening.

The behavior for $\lambda>0$ with $\zeta=1$ is qualitatively different, however. For small positive $\lambda$, $L(t)$ initially exhibits apparent power-law growth, whereas for larger $\lambda$ the domain size eventually saturates. This transition is controlled by the effective coupling parameter $b=2\lambda +\nu-\zeta$, which  measures the strength of active interfacial currents relative to passive interfacial stiffness. Since $b$ is positive  for $\lambda>1/4$ (when $\zeta=1, \nu=1/2$), in this regime we observe strongly suppressed growth. At intermediate positive activity (e.g., $\lambda=1$), the system instead displays an extended transient regime before approaching saturation, as illustrated in Fig.~\ref{fig: AMB$+$_log_correction}(a).

\section{Conclusion and Discussions}
\label{secIV}

In this work, we present a comprehensive numerical study of the critical dynamics and supercritical coarsening in Active Model B (AMB) and its minimal extension, Active Model B$+$ (AMB$+$), focusing on the deterministic limit in two dimensions. Our investigation was motivated by recent functional renormalization group (FRG) predictions~\cite{Fejos2025}, which identified the active couplings $\lambda$, $\nu$, and $\zeta$ as marginally relevant perturbations at the Gaussian fixed point in $d=2$, potentially introducing logarithmic corrections to the classical Lifshitz-Slyozov $t^{1/3}$ coarsening law. Through extensive simulations across a wide range of parameters, we have systematically characterized how these nonequilibrium terms modify phase ordering kinetics and have uncovered a rich interplay between different active contributions that can either enhance or suppress deviations from equilibrium scaling. 
\begin{figure}[t]
    \centering
    \includegraphics[width=\linewidth]{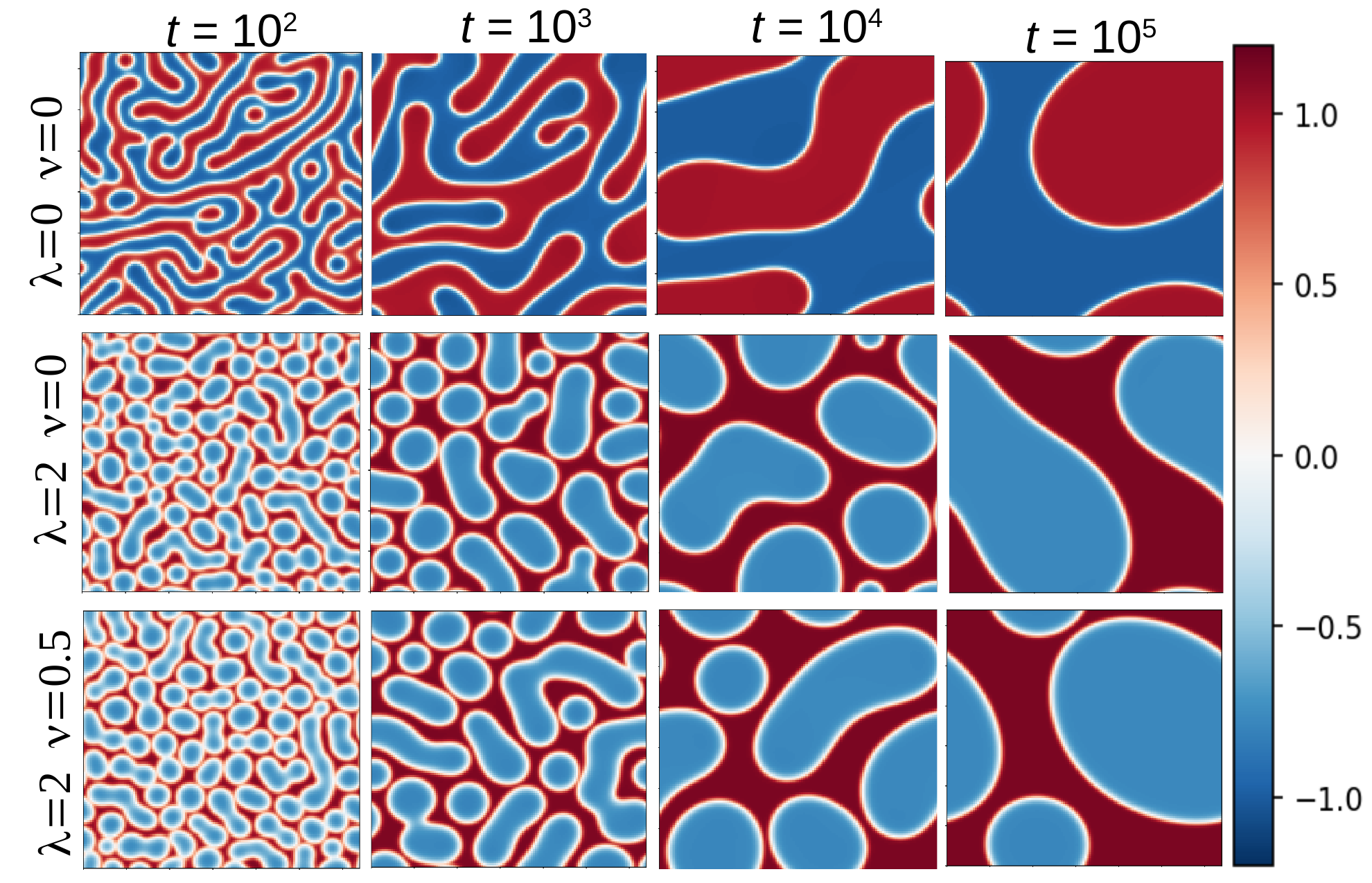}
    \caption{Snapshot of the time evolution of Model B (top), and AMB ($\lambda=2,$ $\nu=0,0.5$) at different time points, starting from a random initial state. The system size considered is $L=128$.}  
    \label{fig: Config}
\end{figure}

At the thermodynamic critical point $r_c=0$, both AMB and AMB$+$ exhibit mean-field critical behavior identical to equilibrium Model B, with exponents $\alpha=1/4$, $z=4$, $\beta=1/2$, and $\nu=1/2$, providing a baseline for future fluctuation-induced effects. We also derived analytical expressions for the pseudodensity and pseudopotential in AMB$+$, validated numerically, yielding a complete phase diagram in the $\phi$-$\lambda$-$r$ plane.

For supercritical quenches, domain growth in AMB and AMB$+$ follows $L(t) \sim t^{1/3}(1+ c/\ln t)$, confirming $\lambda$ as a marginal perturbation that introduces only logarithmic corrections without altering the Lifshitz–Slyozov scaling, precisely matching FRG predictions. While the correction term $c$ is large in AMB, it is suppressed in AMB$+$. In AMB$+$, the $\zeta$ term suppresses the logarithmic deviations further; for $\zeta>0$, the growth law remains close to $t^{1/3}$ with smaller log corrections. For large $|b|$ with $\zeta\lambda<0$, the active current reverses Ostwald ripening, driving the system toward a microphase-separated state where coarsening arrests over extended time windows. Collectively, our results show that marginal activity modulates phase ordering via logarithmic corrections in AMB, while AMB$+$ allows cancellation of the marginal combination through $b = 2\lambda + \nu - \zeta$, enabling a return to pure $t^{1/3}$ scaling or qualitatively new arrested behavior.

The physical origin of this marginality in AMB can be understood from the structure of the active current. The $\lambda$ term generates an interfacial current proportional to the local curvature, with a direction determined by the orientation of the density gradient. Importantly, this contribution does not introduce a new intrinsic length scale; instead, it modifies the efficiency of mass transport along and across interfaces. As a result, early-time coarsening can differ from the equilibrium case, reflecting activity-induced transients, whereas at late times the merging of large domains proceeds similarly to equilibrium Model B dynamics. Figure~\ref{fig: Config} compares the evolution of AMB on a $128\times128$ grid for $\lambda=2$ with $\nu=0$ and $0.5$, alongside the equilibrium case ($\lambda=0$). Despite noticeable differences at short times, the late-stage coarsening morphology and growth dynamics remain qualitatively similar, consistent with the marginal character of the active terms in two dimensions.

Overall, our results provide a unified picture of how marginal activity modifies phase-ordering kinetics in two-dimensional conserved scalar field theories. 
In AMB, the active parameters $\lambda$ and $\nu$ generate logarithmic corrections. In AMB$^+$, additional active terms allow cancellation of the marginal contribution through the effective coupling $b=2\lambda+\nu-\zeta$, resulting in an effective growth exponent close to $1/3$ in appropriate parameter regimes. When this cancellation is incomplete or reversed, active currents can oppose Ostwald ripening and stabilize finite-size domains over extended times.

\begin{acknowledgements} PKM acknowledges the Max Planck Institute for the Physics of Complex Systems  for supporting his visit, where part of this work was completed, and financial support from ANRF$-$SERB, DST, Government of India, under Grant No. MTR/2023/000644. 
\end{acknowledgements}

\appendix
\section{Numerical Simulations}\label{apndx: A}

We integrate the dynamics of AMB$+$ as in Eq.~\eqref{eq: dphidt_AMB$+$} with a pseudo-spectral method employing periodic boundary conditions in $2$D and an Euler-time update scheme. In all our simulations, we set $u=1$, $\kappa=1$, and we take spatial discretization $\Delta x=\Delta y =1$ and time-step $dt=10^{-2}$. We checked that our results are stable upon decreasing $\Delta x$,$\Delta y$, and $dt$.

\textit{Order Parameter $m$}: For a $L\times L$ square gird we take absolute of $\phi(i,j)$ (with $i=j=1,2,\ldots L$) values at each grid point at time $t$ and calculated
\begin{equation*}
    m(\mathbf{r},t) = \frac{1}{L^2}\sum_{i=1}^{L}\sum_{j=1}^{L}|\phi(i,j,t)|,
\end{equation*}
for the estimation of order parameter $m(\mathbf r, t)$ as  in Eq.~\eqref{eq: order_parameter}.

{\it Binodal Density}: 
On a $L\times L$ square grid, we calculate $\phi_\pm$, by measuring all the $\phi(i,j,t)>0$ and $\phi(i,j,t)$ separately at all grid points and at time $t$. Then we calculate the  binodal densities as
\bea 
    \phi_\pm = \frac1{N_\pm}\sum_{i=1}^{L}\sum_{j=1}^{L}
    \phi(i,j,t)\,\Theta(\pm\phi(i,j,t));\cr
    ~~{\rm where} ~~ N_\pm= \sum_{i=1}^{L}\sum_{j=1}^{L} \Theta(\pm\phi(i,j,t))
    \nonumber
\eea
\bibliography{AMBplus.bib}

\begin{thebibliography}{59}%
\makeatletter
\providecommand \@ifxundefined [1]{%
 \@ifx{#1\undefined}
}%
\providecommand \@ifnum [1]{%
 \ifnum #1\expandafter \@firstoftwo
 \else \expandafter \@secondoftwo
 \fi
}%
\providecommand \@ifx [1]{%
 \ifx #1\expandafter \@firstoftwo
 \else \expandafter \@secondoftwo
 \fi
}%
\providecommand \natexlab [1]{#1}%
\providecommand \enquote  [1]{``#1''}%
\providecommand \bibnamefont  [1]{#1}%
\providecommand \bibfnamefont [1]{#1}%
\providecommand \citenamefont [1]{#1}%
\providecommand \href@noop [0]{\@secondoftwo}%
\providecommand \href [0]{\begingroup \@sanitize@url \@href}%
\providecommand \@href[1]{\@@startlink{#1}\@@href}%
\providecommand \@@href[1]{\endgroup#1\@@endlink}%
\providecommand \@sanitize@url [0]{\catcode `\\12\catcode `\$12\catcode `\&12\catcode `\#12\catcode `\^12\catcode `\_12\catcode `\%12\relax}%
\providecommand \@@startlink[1]{}%
\providecommand \@@endlink[0]{}%
\providecommand \url  [0]{\begingroup\@sanitize@url \@url }%
\providecommand \@url [1]{\endgroup\@href {#1}{\urlprefix }}%
\providecommand \urlprefix  [0]{URL }%
\providecommand \Eprint [0]{\href }%
\providecommand \doibase [0]{https://doi.org/}%
\providecommand \selectlanguage [0]{\@gobble}%
\providecommand \bibinfo  [0]{\@secondoftwo}%
\providecommand \bibfield  [0]{\@secondoftwo}%
\providecommand \translation [1]{[#1]}%
\providecommand \BibitemOpen [0]{}%
\providecommand \bibitemStop [0]{}%
\providecommand \bibitemNoStop [0]{.\EOS\space}%
\providecommand \EOS [0]{\spacefactor3000\relax}%
\providecommand \BibitemShut  [1]{\csname bibitem#1\endcsname}%
\let\auto@bib@innerbib\@empty
\bibitem [{\citenamefont {Ramaswamy}(2017)}]{Ramaswamy2017}%
  \BibitemOpen
  \bibfield  {author} {\bibinfo {author} {\bibfnamefont {S.}~\bibnamefont {Ramaswamy}},\ }\bibfield  {title} {\bibinfo {title} {Active matter},\ }\href {https://doi.org/10.1088/1742-5468/aa6bc5} {\bibfield  {journal} {\bibinfo  {journal} {J. Stat. Mech.: Theory Exp.}\ }\textbf {\bibinfo {volume} {2017}}\bibinfo  {number} { (5)},\ \bibinfo {pages} {054002}}\BibitemShut {NoStop}%
\bibitem [{\citenamefont {Marchetti}\ \emph {et~al.}(2013)\citenamefont {Marchetti}, \citenamefont {Joanny}, \citenamefont {Ramaswamy}, \citenamefont {Liverpool}, \citenamefont {Prost}, \citenamefont {Rao},\ and\ \citenamefont {Simha}}]{Marchetti2013}%
  \BibitemOpen
\bibfield  {number} {  }\bibfield  {author} {\bibinfo {author} {\bibfnamefont {M.~C.}\ \bibnamefont {Marchetti}}, \bibinfo {author} {\bibfnamefont {J.~F.}\ \bibnamefont {Joanny}}, \bibinfo {author} {\bibfnamefont {S.}~\bibnamefont {Ramaswamy}}, \bibinfo {author} {\bibfnamefont {T.~B.}\ \bibnamefont {Liverpool}}, \bibinfo {author} {\bibfnamefont {J.}~\bibnamefont {Prost}}, \bibinfo {author} {\bibfnamefont {M.}~\bibnamefont {Rao}},\ and\ \bibinfo {author} {\bibfnamefont {R.~A.}\ \bibnamefont {Simha}},\ }\bibfield  {title} {\bibinfo {title} {Hydrodynamics of soft active matter},\ }\href {https://doi.org/10.1103/revmodphys.85.1143} {\bibfield  {journal} {\bibinfo  {journal} {Rev. Mod. Phys.}\ }\textbf {\bibinfo {volume} {85}},\ \bibinfo {pages} {1143} (\bibinfo {year} {2013})}\BibitemShut {NoStop}%
\bibitem [{\citenamefont {Katz}\ \emph {et~al.}(2011)\citenamefont {Katz}, \citenamefont {Tunstrøm}, \citenamefont {Ioannou}, \citenamefont {Huepe},\ and\ \citenamefont {Couzin}}]{Katz2011}%
  \BibitemOpen
  \bibfield  {author} {\bibinfo {author} {\bibfnamefont {Y.}~\bibnamefont {Katz}}, \bibinfo {author} {\bibfnamefont {K.}~\bibnamefont {Tunstrøm}}, \bibinfo {author} {\bibfnamefont {C.~C.}\ \bibnamefont {Ioannou}}, \bibinfo {author} {\bibfnamefont {C.}~\bibnamefont {Huepe}},\ and\ \bibinfo {author} {\bibfnamefont {I.~D.}\ \bibnamefont {Couzin}},\ }\bibfield  {title} {\bibinfo {title} {Inferring the structure and dynamics of interactions in schooling fish},\ }\href {https://doi.org/10.1073/pnas.1107583108} {\bibfield  {journal} {\bibinfo  {journal} {Proc. Natl. Acad. Sci.}\ }\textbf {\bibinfo {volume} {108}},\ \bibinfo {pages} {18720} (\bibinfo {year} {2011})}\BibitemShut {NoStop}%
\bibitem [{\citenamefont {Ballerini}\ \emph {et~al.}(2008)\citenamefont {Ballerini}, \citenamefont {Cabibbo}, \citenamefont {Candelier}, \citenamefont {Cavagna}, \citenamefont {Cisbani}, \citenamefont {Giardina}, \citenamefont {Lecomte}, \citenamefont {Orlandi}, \citenamefont {Parisi}, \citenamefont {Procaccini}, \citenamefont {Viale},\ and\ \citenamefont {Zdravkovic}}]{Ballerini2008}%
  \BibitemOpen
  \bibfield  {author} {\bibinfo {author} {\bibfnamefont {M.}~\bibnamefont {Ballerini}}, \bibinfo {author} {\bibfnamefont {N.}~\bibnamefont {Cabibbo}}, \bibinfo {author} {\bibfnamefont {R.}~\bibnamefont {Candelier}}, \bibinfo {author} {\bibfnamefont {A.}~\bibnamefont {Cavagna}}, \bibinfo {author} {\bibfnamefont {E.}~\bibnamefont {Cisbani}}, \bibinfo {author} {\bibfnamefont {I.}~\bibnamefont {Giardina}}, \bibinfo {author} {\bibfnamefont {V.}~\bibnamefont {Lecomte}}, \bibinfo {author} {\bibfnamefont {A.}~\bibnamefont {Orlandi}}, \bibinfo {author} {\bibfnamefont {G.}~\bibnamefont {Parisi}}, \bibinfo {author} {\bibfnamefont {A.}~\bibnamefont {Procaccini}}, \bibinfo {author} {\bibfnamefont {M.}~\bibnamefont {Viale}},\ and\ \bibinfo {author} {\bibfnamefont {V.}~\bibnamefont {Zdravkovic}},\ }\bibfield  {title} {\bibinfo {title} {Interaction ruling animal collective behavior depends on topological rather than metric distance: Evidence from a field study},\ }\href {https://doi.org/10.1073/pnas.0711437105} {\bibfield
  {journal} {\bibinfo  {journal} {Proc. Natl. Acad. Sci.}\ }\textbf {\bibinfo {volume} {105}},\ \bibinfo {pages} {1232} (\bibinfo {year} {2008})}\BibitemShut {NoStop}%
\bibitem [{\citenamefont {Marchetti}\ \emph {et~al.}(2016)\citenamefont {Marchetti}, \citenamefont {Fily}, \citenamefont {Henkes}, \citenamefont {Patch},\ and\ \citenamefont {Yllanes}}]{Marchetti2016}%
  \BibitemOpen
  \bibfield  {author} {\bibinfo {author} {\bibfnamefont {M.~C.}\ \bibnamefont {Marchetti}}, \bibinfo {author} {\bibfnamefont {Y.}~\bibnamefont {Fily}}, \bibinfo {author} {\bibfnamefont {S.}~\bibnamefont {Henkes}}, \bibinfo {author} {\bibfnamefont {A.}~\bibnamefont {Patch}},\ and\ \bibinfo {author} {\bibfnamefont {D.}~\bibnamefont {Yllanes}},\ }\bibfield  {title} {\bibinfo {title} {Minimal model of active colloids highlights the role of mechanical interactions in controlling the emergent behavior of active matter},\ }\href {https://doi.org/10.1016/j.cocis.2016.01.003} {\bibfield  {journal} {\bibinfo  {journal} {Curr. Opin. Colloid \& Interface Sci.}\ }\textbf {\bibinfo {volume} {21}},\ \bibinfo {pages} {34} (\bibinfo {year} {2016})}\BibitemShut {NoStop}%
\bibitem [{\citenamefont {Farage}\ \emph {et~al.}(2015)\citenamefont {Farage}, \citenamefont {Krinninger},\ and\ \citenamefont {Brader}}]{Farage2015}%
  \BibitemOpen
  \bibfield  {author} {\bibinfo {author} {\bibfnamefont {T.~F.~F.}\ \bibnamefont {Farage}}, \bibinfo {author} {\bibfnamefont {P.}~\bibnamefont {Krinninger}},\ and\ \bibinfo {author} {\bibfnamefont {J.~M.}\ \bibnamefont {Brader}},\ }\bibfield  {title} {\bibinfo {title} {Effective interactions in active brownian suspensions},\ }\href {https://doi.org/10.1103/PhysRevE.91.042310} {\bibfield  {journal} {\bibinfo  {journal} {Phys. Rev. E}\ }\textbf {\bibinfo {volume} {91}},\ \bibinfo {pages} {042310} (\bibinfo {year} {2015})}\BibitemShut {NoStop}%
\bibitem [{\citenamefont {Liebchen}\ \emph {et~al.}(2017)\citenamefont {Liebchen}, \citenamefont {Marenduzzo},\ and\ \citenamefont {Cates}}]{Liebchen2017}%
  \BibitemOpen
  \bibfield  {author} {\bibinfo {author} {\bibfnamefont {B.}~\bibnamefont {Liebchen}}, \bibinfo {author} {\bibfnamefont {D.}~\bibnamefont {Marenduzzo}},\ and\ \bibinfo {author} {\bibfnamefont {M.~E.}\ \bibnamefont {Cates}},\ }\bibfield  {title} {\bibinfo {title} {Phoretic interactions generically induce dynamic clusters and wave patterns in active colloids},\ }\href {https://doi.org/10.1103/PhysRevLett.118.268001} {\bibfield  {journal} {\bibinfo  {journal} {Phys. Rev. Lett.}\ }\textbf {\bibinfo {volume} {118}},\ \bibinfo {pages} {268001} (\bibinfo {year} {2017})}\BibitemShut {NoStop}%
\bibitem [{\citenamefont {Illien}\ \emph {et~al.}(2017)\citenamefont {Illien}, \citenamefont {Golestanian},\ and\ \citenamefont {Sen}}]{Illien2017}%
  \BibitemOpen
  \bibfield  {author} {\bibinfo {author} {\bibfnamefont {P.}~\bibnamefont {Illien}}, \bibinfo {author} {\bibfnamefont {R.}~\bibnamefont {Golestanian}},\ and\ \bibinfo {author} {\bibfnamefont {A.}~\bibnamefont {Sen}},\ }\bibfield  {title} {\bibinfo {title} {'fuelled' motion: phoretic motility and collective behaviour of active colloids},\ }\href {https://doi.org/10.1039/c7cs00087a} {\bibfield  {journal} {\bibinfo  {journal} {Chem. Soc. Rev.}\ }\textbf {\bibinfo {volume} {46}},\ \bibinfo {pages} {5508} (\bibinfo {year} {2017})}\BibitemShut {NoStop}%
\bibitem [{\citenamefont {van~der Linden}\ \emph {et~al.}(2019)\citenamefont {van~der Linden}, \citenamefont {Alexander}, \citenamefont {Aarts},\ and\ \citenamefont {Dauchot}}]{vanderLinden2019}%
  \BibitemOpen
  \bibfield  {author} {\bibinfo {author} {\bibfnamefont {M.~N.}\ \bibnamefont {van~der Linden}}, \bibinfo {author} {\bibfnamefont {L.~C.}\ \bibnamefont {Alexander}}, \bibinfo {author} {\bibfnamefont {D.~G. A.~L.}\ \bibnamefont {Aarts}},\ and\ \bibinfo {author} {\bibfnamefont {O.}~\bibnamefont {Dauchot}},\ }\bibfield  {title} {\bibinfo {title} {Interrupted motility induced phase separation in aligning active colloids},\ }\href {https://doi.org/10.1103/PhysRevLett.123.098001} {\bibfield  {journal} {\bibinfo  {journal} {Phys. Rev. Lett.}\ }\textbf {\bibinfo {volume} {123}},\ \bibinfo {pages} {098001} (\bibinfo {year} {2019})}\BibitemShut {NoStop}%
\bibitem [{\citenamefont {Z\"{o}ttl}\ and\ \citenamefont {Stark}(2023)}]{Zttl2023}%
  \BibitemOpen
  \bibfield  {author} {\bibinfo {author} {\bibfnamefont {A.}~\bibnamefont {Z\"{o}ttl}}\ and\ \bibinfo {author} {\bibfnamefont {H.}~\bibnamefont {Stark}},\ }\bibfield  {title} {\bibinfo {title} {Modeling active colloids: From active brownian particles to hydrodynamic and chemical fields},\ }\href {https://doi.org/10.1146/annurev-conmatphys-040821-115500} {\bibfield  {journal} {\bibinfo  {journal} {Annu. Rev. Condens. Matter Phys.}\ }\textbf {\bibinfo {volume} {14}},\ \bibinfo {pages} {109} (\bibinfo {year} {2023})}\BibitemShut {NoStop}%
\bibitem [{\citenamefont {Melio}\ \emph {et~al.}(2025)\citenamefont {Melio}, \citenamefont {Riedel}, \citenamefont {Azadbakht}, \citenamefont {Caipa~Cure}, \citenamefont {Evers}, \citenamefont {Babaei}, \citenamefont {Mashaghi}, \citenamefont {de~Graaf},\ and\ \citenamefont {Kraft}}]{Melio2025}%
  \BibitemOpen
  \bibfield  {author} {\bibinfo {author} {\bibfnamefont {J.}~\bibnamefont {Melio}}, \bibinfo {author} {\bibfnamefont {S.}~\bibnamefont {Riedel}}, \bibinfo {author} {\bibfnamefont {A.}~\bibnamefont {Azadbakht}}, \bibinfo {author} {\bibfnamefont {S.~A.}\ \bibnamefont {Caipa~Cure}}, \bibinfo {author} {\bibfnamefont {T.~M.}\ \bibnamefont {Evers}}, \bibinfo {author} {\bibfnamefont {M.}~\bibnamefont {Babaei}}, \bibinfo {author} {\bibfnamefont {A.}~\bibnamefont {Mashaghi}}, \bibinfo {author} {\bibfnamefont {J.}~\bibnamefont {de~Graaf}},\ and\ \bibinfo {author} {\bibfnamefont {D.~J.}\ \bibnamefont {Kraft}},\ }\bibfield  {title} {\bibinfo {title} {The motion of catalytically active colloids approaching a surface},\ }\href {https://doi.org/10.1039/d4sm01387e} {\bibfield  {journal} {\bibinfo  {journal} {Soft Matter}\ }\textbf {\bibinfo {volume} {21}},\ \bibinfo {pages} {2541} (\bibinfo {year} {2025})}\BibitemShut {NoStop}%
\bibitem [{\citenamefont {Bendix}\ \emph {et~al.}(2008)\citenamefont {Bendix}, \citenamefont {Koenderink}, \citenamefont {Cuvelier}, \citenamefont {Dogic}, \citenamefont {Koeleman}, \citenamefont {Brieher}, \citenamefont {Field}, \citenamefont {Mahadevan},\ and\ \citenamefont {Weitz}}]{Bendix2008}%
  \BibitemOpen
  \bibfield  {author} {\bibinfo {author} {\bibfnamefont {P.~M.}\ \bibnamefont {Bendix}}, \bibinfo {author} {\bibfnamefont {G.~H.}\ \bibnamefont {Koenderink}}, \bibinfo {author} {\bibfnamefont {D.}~\bibnamefont {Cuvelier}}, \bibinfo {author} {\bibfnamefont {Z.}~\bibnamefont {Dogic}}, \bibinfo {author} {\bibfnamefont {B.~N.}\ \bibnamefont {Koeleman}}, \bibinfo {author} {\bibfnamefont {W.~M.}\ \bibnamefont {Brieher}}, \bibinfo {author} {\bibfnamefont {C.~M.}\ \bibnamefont {Field}}, \bibinfo {author} {\bibfnamefont {L.}~\bibnamefont {Mahadevan}},\ and\ \bibinfo {author} {\bibfnamefont {D.~A.}\ \bibnamefont {Weitz}},\ }\bibfield  {title} {\bibinfo {title} {A quantitative analysis of contractility in active cytoskeletal protein networks},\ }\href {https://doi.org/10.1529/biophysj.107.117960} {\bibfield  {journal} {\bibinfo  {journal} {Biophys. J.}\ }\textbf {\bibinfo {volume} {94}},\ \bibinfo {pages} {3126} (\bibinfo {year} {2008})}\BibitemShut {NoStop}%
\bibitem [{\citenamefont {Sanchez}\ \emph {et~al.}(2012)\citenamefont {Sanchez}, \citenamefont {Chen}, \citenamefont {DeCamp}, \citenamefont {Heymann},\ and\ \citenamefont {Dogic}}]{Sanchez2012}%
  \BibitemOpen
  \bibfield  {author} {\bibinfo {author} {\bibfnamefont {T.}~\bibnamefont {Sanchez}}, \bibinfo {author} {\bibfnamefont {D.~T.~N.}\ \bibnamefont {Chen}}, \bibinfo {author} {\bibfnamefont {S.~J.}\ \bibnamefont {DeCamp}}, \bibinfo {author} {\bibfnamefont {M.}~\bibnamefont {Heymann}},\ and\ \bibinfo {author} {\bibfnamefont {Z.}~\bibnamefont {Dogic}},\ }\bibfield  {title} {\bibinfo {title} {Spontaneous motion in hierarchically assembled active matter},\ }\href {https://doi.org/10.1038/nature11591} {\bibfield  {journal} {\bibinfo  {journal} {Nature}\ }\textbf {\bibinfo {volume} {491}},\ \bibinfo {pages} {431} (\bibinfo {year} {2012})}\BibitemShut {NoStop}%
\bibitem [{\citenamefont {Alvarado}\ \emph {et~al.}(2013)\citenamefont {Alvarado}, \citenamefont {Sheinman}, \citenamefont {Sharma}, \citenamefont {MacKintosh},\ and\ \citenamefont {Koenderink}}]{Alvarado2013}%
  \BibitemOpen
  \bibfield  {author} {\bibinfo {author} {\bibfnamefont {J.}~\bibnamefont {Alvarado}}, \bibinfo {author} {\bibfnamefont {M.}~\bibnamefont {Sheinman}}, \bibinfo {author} {\bibfnamefont {A.}~\bibnamefont {Sharma}}, \bibinfo {author} {\bibfnamefont {F.~C.}\ \bibnamefont {MacKintosh}},\ and\ \bibinfo {author} {\bibfnamefont {G.~H.}\ \bibnamefont {Koenderink}},\ }\bibfield  {title} {\bibinfo {title} {Molecular motors robustly drive active gels to a critically connected state},\ }\href {https://doi.org/10.1038/nphys2715} {\bibfield  {journal} {\bibinfo  {journal} {Nat. Phys.}\ }\textbf {\bibinfo {volume} {9}},\ \bibinfo {pages} {591} (\bibinfo {year} {2013})}\BibitemShut {NoStop}%
\bibitem [{\citenamefont {Geisterfer}\ \emph {et~al.}(2021)\citenamefont {Geisterfer}, \citenamefont {Guilloux}, \citenamefont {Gatlin},\ and\ \citenamefont {Gibeaux}}]{Geisterfer2021}%
  \BibitemOpen
  \bibfield  {author} {\bibinfo {author} {\bibfnamefont {Z.~M.}\ \bibnamefont {Geisterfer}}, \bibinfo {author} {\bibfnamefont {G.}~\bibnamefont {Guilloux}}, \bibinfo {author} {\bibfnamefont {J.~C.}\ \bibnamefont {Gatlin}},\ and\ \bibinfo {author} {\bibfnamefont {R.}~\bibnamefont {Gibeaux}},\ }\bibfield  {title} {\bibinfo {title} {The cytoskeleton and its roles in self-organization phenomena: Insights from xenopus egg extracts},\ }\href {https://doi.org/10.3390/cells10092197} {\bibfield  {journal} {\bibinfo  {journal} {Cells}\ }\textbf {\bibinfo {volume} {10}},\ \bibinfo {pages} {2197} (\bibinfo {year} {2021})}\BibitemShut {NoStop}%
\bibitem [{\citenamefont {J\"{u}licher}\ \emph {et~al.}(1997)\citenamefont {J\"{u}licher}, \citenamefont {Ajdari},\ and\ \citenamefont {Prost}}]{Jlicher1997}%
  \BibitemOpen
  \bibfield  {author} {\bibinfo {author} {\bibfnamefont {F.}~\bibnamefont {J\"{u}licher}}, \bibinfo {author} {\bibfnamefont {A.}~\bibnamefont {Ajdari}},\ and\ \bibinfo {author} {\bibfnamefont {J.}~\bibnamefont {Prost}},\ }\bibfield  {title} {\bibinfo {title} {Modeling molecular motors},\ }\href {https://doi.org/10.1103/revmodphys.69.1269} {\bibfield  {journal} {\bibinfo  {journal} {Rev. Mod. Phys.}\ }\textbf {\bibinfo {volume} {69}},\ \bibinfo {pages} {1269} (\bibinfo {year} {1997})}\BibitemShut {NoStop}%
\bibitem [{\citenamefont {Ndlec}\ \emph {et~al.}(1997)\citenamefont {Ndlec}, \citenamefont {Surrey}, \citenamefont {Maggs},\ and\ \citenamefont {Leibler}}]{Ndlec1997}%
  \BibitemOpen
  \bibfield  {author} {\bibinfo {author} {\bibfnamefont {F.~J.}\ \bibnamefont {Ndlec}}, \bibinfo {author} {\bibfnamefont {T.}~\bibnamefont {Surrey}}, \bibinfo {author} {\bibfnamefont {A.~C.}\ \bibnamefont {Maggs}},\ and\ \bibinfo {author} {\bibfnamefont {S.}~\bibnamefont {Leibler}},\ }\bibfield  {title} {\bibinfo {title} {Self-organization of microtubules and motors},\ }\href {https://doi.org/10.1038/38532} {\bibfield  {journal} {\bibinfo  {journal} {Nature}\ }\textbf {\bibinfo {volume} {389}},\ \bibinfo {pages} {305} (\bibinfo {year} {1997})}\BibitemShut {NoStop}%
\bibitem [{\citenamefont {Kruse}\ \emph {et~al.}(2004)\citenamefont {Kruse}, \citenamefont {Joanny}, \citenamefont {J\"{u}licher}, \citenamefont {Prost},\ and\ \citenamefont {Sekimoto}}]{Kruse2004}%
  \BibitemOpen
  \bibfield  {author} {\bibinfo {author} {\bibfnamefont {K.}~\bibnamefont {Kruse}}, \bibinfo {author} {\bibfnamefont {J.~F.}\ \bibnamefont {Joanny}}, \bibinfo {author} {\bibfnamefont {F.}~\bibnamefont {J\"{u}licher}}, \bibinfo {author} {\bibfnamefont {J.}~\bibnamefont {Prost}},\ and\ \bibinfo {author} {\bibfnamefont {K.}~\bibnamefont {Sekimoto}},\ }\bibfield  {title} {\bibinfo {title} {Asters, vortices, and rotating spirals in active gels of polar filaments},\ }\href {https://doi.org/10.1103/physrevlett.92.078101} {\bibfield  {journal} {\bibinfo  {journal} {Phys. Rev. Lett.}\ }\textbf {\bibinfo {volume} {92}},\ \bibinfo {pages} {078101} (\bibinfo {year} {2004})}\BibitemShut {NoStop}%
\bibitem [{\citenamefont {Liverpool}\ and\ \citenamefont {Marchetti}(2003)}]{Liverpool2003}%
  \BibitemOpen
  \bibfield  {author} {\bibinfo {author} {\bibfnamefont {T.~B.}\ \bibnamefont {Liverpool}}\ and\ \bibinfo {author} {\bibfnamefont {M.~C.}\ \bibnamefont {Marchetti}},\ }\bibfield  {title} {\bibinfo {title} {Instabilities of isotropic solutions of active polar filaments},\ }\href {https://doi.org/10.1103/physrevlett.90.138102} {\bibfield  {journal} {\bibinfo  {journal} {Phys. Rev. Lett.}\ }\textbf {\bibinfo {volume} {90}},\ \bibinfo {pages} {138102} (\bibinfo {year} {2003})}\BibitemShut {NoStop}%
\bibitem [{\citenamefont {Liverpool}\ and\ \citenamefont {Marchetti}(2008)}]{Liverpool2008}%
  \BibitemOpen
  \bibfield  {author} {\bibinfo {author} {\bibfnamefont {T.~B.}\ \bibnamefont {Liverpool}}\ and\ \bibinfo {author} {\bibfnamefont {M.~C.}\ \bibnamefont {Marchetti}},\ }\bibinfo {title} {Hydrodynamics and rheology of active polar filaments},\ in\ \href {https://doi.org/10.1007/978-0-387-73050-9_7} {\emph {\bibinfo {booktitle} {Cell Motility}}}\ (\bibinfo  {publisher} {Springer New York},\ \bibinfo {year} {2008})\ p.\ \bibinfo {pages} {177–206}\BibitemShut {NoStop}%
\bibitem [{\citenamefont {Schaller}\ \emph {et~al.}(2010)\citenamefont {Schaller}, \citenamefont {Weber}, \citenamefont {Semmrich}, \citenamefont {Frey},\ and\ \citenamefont {Bausch}}]{Schaller2010}%
  \BibitemOpen
  \bibfield  {author} {\bibinfo {author} {\bibfnamefont {V.}~\bibnamefont {Schaller}}, \bibinfo {author} {\bibfnamefont {C.}~\bibnamefont {Weber}}, \bibinfo {author} {\bibfnamefont {C.}~\bibnamefont {Semmrich}}, \bibinfo {author} {\bibfnamefont {E.}~\bibnamefont {Frey}},\ and\ \bibinfo {author} {\bibfnamefont {A.~R.}\ \bibnamefont {Bausch}},\ }\bibfield  {title} {\bibinfo {title} {Polar patterns of driven filaments},\ }\href {https://doi.org/10.1038/nature09312} {\bibfield  {journal} {\bibinfo  {journal} {Nature}\ }\textbf {\bibinfo {volume} {467}},\ \bibinfo {pages} {73} (\bibinfo {year} {2010})}\BibitemShut {NoStop}%
\bibitem [{\citenamefont {Tailleur}\ and\ \citenamefont {Cates}(2008)}]{Tailleur2008}%
  \BibitemOpen
  \bibfield  {author} {\bibinfo {author} {\bibfnamefont {J.}~\bibnamefont {Tailleur}}\ and\ \bibinfo {author} {\bibfnamefont {M.~E.}\ \bibnamefont {Cates}},\ }\bibfield  {title} {\bibinfo {title} {Statistical mechanics of interacting run-and-tumble bacteria},\ }\href {https://doi.org/10.1103/physrevlett.100.218103} {\bibfield  {journal} {\bibinfo  {journal} {Phys. Rev. Lett.}\ }\textbf {\bibinfo {volume} {100}},\ \bibinfo {pages} {218103} (\bibinfo {year} {2008})}\BibitemShut {NoStop}%
\bibitem [{\citenamefont {Cates}\ and\ \citenamefont {Tailleur}(2015)}]{Cates2015}%
  \BibitemOpen
  \bibfield  {author} {\bibinfo {author} {\bibfnamefont {M.~E.}\ \bibnamefont {Cates}}\ and\ \bibinfo {author} {\bibfnamefont {J.}~\bibnamefont {Tailleur}},\ }\bibfield  {title} {\bibinfo {title} {Motility-induced phase separation},\ }\href {https://doi.org/10.1146/annurev-conmatphys-031214-014710} {\bibfield  {journal} {\bibinfo  {journal} {Annu. Rev. Condens. Matter Phys.}\ }\textbf {\bibinfo {volume} {6}},\ \bibinfo {pages} {219} (\bibinfo {year} {2015})}\BibitemShut {NoStop}%
\bibitem [{\citenamefont {Fily}\ and\ \citenamefont {Marchetti}(2012)}]{Fily2012}%
  \BibitemOpen
  \bibfield  {author} {\bibinfo {author} {\bibfnamefont {Y.}~\bibnamefont {Fily}}\ and\ \bibinfo {author} {\bibfnamefont {M.~C.}\ \bibnamefont {Marchetti}},\ }\bibfield  {title} {\bibinfo {title} {Athermal phase separation of self-propelled particles with no alignment},\ }\href {https://doi.org/10.1103/physrevlett.108.235702} {\bibfield  {journal} {\bibinfo  {journal} {Phys. Rev. Lett.}\ }\textbf {\bibinfo {volume} {108}},\ \bibinfo {pages} {235702} (\bibinfo {year} {2012})}\BibitemShut {NoStop}%
\bibitem [{\citenamefont {Redner}\ \emph {et~al.}(2013)\citenamefont {Redner}, \citenamefont {Hagan},\ and\ \citenamefont {Baskaran}}]{Redner2013}%
  \BibitemOpen
  \bibfield  {author} {\bibinfo {author} {\bibfnamefont {G.~S.}\ \bibnamefont {Redner}}, \bibinfo {author} {\bibfnamefont {M.~F.}\ \bibnamefont {Hagan}},\ and\ \bibinfo {author} {\bibfnamefont {A.}~\bibnamefont {Baskaran}},\ }\bibfield  {title} {\bibinfo {title} {Structure and dynamics of a phase-separating active colloidal fluid},\ }\href {https://doi.org/10.1103/physrevlett.110.055701} {\bibfield  {journal} {\bibinfo  {journal} {Phys. Rev. Lett.}\ }\textbf {\bibinfo {volume} {110}},\ \bibinfo {pages} {055701} (\bibinfo {year} {2013})}\BibitemShut {NoStop}%
\bibitem [{\citenamefont {Stenhammar}\ \emph {et~al.}(2013)\citenamefont {Stenhammar}, \citenamefont {Tiribocchi}, \citenamefont {Allen}, \citenamefont {Marenduzzo},\ and\ \citenamefont {Cates}}]{Stenhammar2013}%
  \BibitemOpen
  \bibfield  {author} {\bibinfo {author} {\bibfnamefont {J.}~\bibnamefont {Stenhammar}}, \bibinfo {author} {\bibfnamefont {A.}~\bibnamefont {Tiribocchi}}, \bibinfo {author} {\bibfnamefont {R.~J.}\ \bibnamefont {Allen}}, \bibinfo {author} {\bibfnamefont {D.}~\bibnamefont {Marenduzzo}},\ and\ \bibinfo {author} {\bibfnamefont {M.~E.}\ \bibnamefont {Cates}},\ }\bibfield  {title} {\bibinfo {title} {Continuum theory of phase separation kinetics for active brownian particles},\ }\href {https://doi.org/10.1103/physrevlett.111.145702} {\bibfield  {journal} {\bibinfo  {journal} {Phys. Rev. Lett.}\ }\textbf {\bibinfo {volume} {111}},\ \bibinfo {pages} {145702} (\bibinfo {year} {2013})}\BibitemShut {NoStop}%
\bibitem [{\citenamefont {Stenhammar}\ \emph {et~al.}(2014)\citenamefont {Stenhammar}, \citenamefont {Marenduzzo}, \citenamefont {Allen},\ and\ \citenamefont {Cates}}]{Stenhammar2014}%
  \BibitemOpen
  \bibfield  {author} {\bibinfo {author} {\bibfnamefont {J.}~\bibnamefont {Stenhammar}}, \bibinfo {author} {\bibfnamefont {D.}~\bibnamefont {Marenduzzo}}, \bibinfo {author} {\bibfnamefont {R.~J.}\ \bibnamefont {Allen}},\ and\ \bibinfo {author} {\bibfnamefont {M.~E.}\ \bibnamefont {Cates}},\ }\bibfield  {title} {\bibinfo {title} {Phase behaviour of active brownian particles: the role of dimensionality},\ }\href {https://doi.org/10.1039/c3sm52813h} {\bibfield  {journal} {\bibinfo  {journal} {Soft Matter}\ }\textbf {\bibinfo {volume} {10}},\ \bibinfo {pages} {1489} (\bibinfo {year} {2014})}\BibitemShut {NoStop}%
\bibitem [{\citenamefont {Bhowmick}\ \emph {et~al.}(2025)\citenamefont {Bhowmick}, \citenamefont {Mitra},\ and\ \citenamefont {Mohanty}}]{Bhowmick2025}%
  \BibitemOpen
  \bibfield  {author} {\bibinfo {author} {\bibfnamefont {A.}~\bibnamefont {Bhowmick}}, \bibinfo {author} {\bibfnamefont {S.}~\bibnamefont {Mitra}},\ and\ \bibinfo {author} {\bibfnamefont {P.~K.}\ \bibnamefont {Mohanty}},\ }\bibfield  {title} {\bibinfo {title} {Geometric and nonequilibrium criticality in run-and-tumble particles with competing motility and attraction},\ }\href {https://doi.org/10.1103/nw6q-868p} {\bibfield  {journal} {\bibinfo  {journal} {Phys. Rev. E}\ }\textbf {\bibinfo {volume} {112}},\ \bibinfo {pages} {044129} (\bibinfo {year} {2025})}\BibitemShut {NoStop}%
\bibitem [{\citenamefont {Buttinoni}\ \emph {et~al.}(2013)\citenamefont {Buttinoni}, \citenamefont {Bialké}, \citenamefont {K\"{u}mmel}, \citenamefont {L\"{o}wen}, \citenamefont {Bechinger},\ and\ \citenamefont {Speck}}]{Buttinoni2013}%
  \BibitemOpen
  \bibfield  {author} {\bibinfo {author} {\bibfnamefont {I.}~\bibnamefont {Buttinoni}}, \bibinfo {author} {\bibfnamefont {J.}~\bibnamefont {Bialké}}, \bibinfo {author} {\bibfnamefont {F.}~\bibnamefont {K\"{u}mmel}}, \bibinfo {author} {\bibfnamefont {H.}~\bibnamefont {L\"{o}wen}}, \bibinfo {author} {\bibfnamefont {C.}~\bibnamefont {Bechinger}},\ and\ \bibinfo {author} {\bibfnamefont {T.}~\bibnamefont {Speck}},\ }\bibfield  {title} {\bibinfo {title} {Dynamical clustering and phase separation in suspensions of self-propelled colloidal particles},\ }\href {https://doi.org/10.1103/physrevlett.110.238301} {\bibfield  {journal} {\bibinfo  {journal} {Phys. Rev. Lett.}\ }\textbf {\bibinfo {volume} {110}},\ \bibinfo {pages} {238301} (\bibinfo {year} {2013})}\BibitemShut {NoStop}%
\bibitem [{\citenamefont {Liu}\ \emph {et~al.}(2019)\citenamefont {Liu}, \citenamefont {Patch}, \citenamefont {Bahar}, \citenamefont {Yllanes}, \citenamefont {Welch}, \citenamefont {Marchetti}, \citenamefont {Thutupalli},\ and\ \citenamefont {Shaevitz}}]{Liu2019}%
  \BibitemOpen
  \bibfield  {author} {\bibinfo {author} {\bibfnamefont {G.}~\bibnamefont {Liu}}, \bibinfo {author} {\bibfnamefont {A.}~\bibnamefont {Patch}}, \bibinfo {author} {\bibfnamefont {F.}~\bibnamefont {Bahar}}, \bibinfo {author} {\bibfnamefont {D.}~\bibnamefont {Yllanes}}, \bibinfo {author} {\bibfnamefont {R.~D.}\ \bibnamefont {Welch}}, \bibinfo {author} {\bibfnamefont {M.~C.}\ \bibnamefont {Marchetti}}, \bibinfo {author} {\bibfnamefont {S.}~\bibnamefont {Thutupalli}},\ and\ \bibinfo {author} {\bibfnamefont {J.~W.}\ \bibnamefont {Shaevitz}},\ }\bibfield  {title} {\bibinfo {title} {Self-driven phase transitions drivemyxococcus xanthusfruiting body formation},\ }\href {https://doi.org/10.1103/physrevlett.122.248102} {\bibfield  {journal} {\bibinfo  {journal} {Phys. Rev. Lett.}\ }\textbf {\bibinfo {volume} {122}},\ \bibinfo {pages} {248102} (\bibinfo {year} {2019})}\BibitemShut {NoStop}%
\bibitem [{\citenamefont {Shaebani}\ \emph {et~al.}(2020)\citenamefont {Shaebani}, \citenamefont {Wysocki}, \citenamefont {Winkler}, \citenamefont {Gompper},\ and\ \citenamefont {Rieger}}]{Shaebani2020}%
  \BibitemOpen
  \bibfield  {author} {\bibinfo {author} {\bibfnamefont {M.~R.}\ \bibnamefont {Shaebani}}, \bibinfo {author} {\bibfnamefont {A.}~\bibnamefont {Wysocki}}, \bibinfo {author} {\bibfnamefont {R.~G.}\ \bibnamefont {Winkler}}, \bibinfo {author} {\bibfnamefont {G.}~\bibnamefont {Gompper}},\ and\ \bibinfo {author} {\bibfnamefont {H.}~\bibnamefont {Rieger}},\ }\bibfield  {title} {\bibinfo {title} {Computational models for active matter},\ }\href {https://doi.org/10.1038/s42254-020-0152-1} {\bibfield  {journal} {\bibinfo  {journal} {Nat. Rev. Phys.}\ }\textbf {\bibinfo {volume} {2}},\ \bibinfo {pages} {181} (\bibinfo {year} {2020})}\BibitemShut {NoStop}%
\bibitem [{\citenamefont {Solon}\ \emph {et~al.}(2015)\citenamefont {Solon}, \citenamefont {Cates},\ and\ \citenamefont {Tailleur}}]{Solon2015}%
  \BibitemOpen
  \bibfield  {author} {\bibinfo {author} {\bibfnamefont {A.~P.}\ \bibnamefont {Solon}}, \bibinfo {author} {\bibfnamefont {M.~E.}\ \bibnamefont {Cates}},\ and\ \bibinfo {author} {\bibfnamefont {J.}~\bibnamefont {Tailleur}},\ }\bibfield  {title} {\bibinfo {title} {Active brownian particles and run-and-tumble particles: A comparative study},\ }\href {https://doi.org/10.1140/epjst/e2015-02457-0} {\bibfield  {journal} {\bibinfo  {journal} {Eur. Phys. J. Spec. Top.}\ }\textbf {\bibinfo {volume} {224}},\ \bibinfo {pages} {1231} (\bibinfo {year} {2015})}\BibitemShut {NoStop}%
\bibitem [{\citenamefont {Speck}\ \emph {et~al.}(2014)\citenamefont {Speck}, \citenamefont {Bialké}, \citenamefont {Menzel},\ and\ \citenamefont {L\"{o}wen}}]{Speck2014}%
  \BibitemOpen
  \bibfield  {author} {\bibinfo {author} {\bibfnamefont {T.}~\bibnamefont {Speck}}, \bibinfo {author} {\bibfnamefont {J.}~\bibnamefont {Bialké}}, \bibinfo {author} {\bibfnamefont {A.~M.}\ \bibnamefont {Menzel}},\ and\ \bibinfo {author} {\bibfnamefont {H.}~\bibnamefont {L\"{o}wen}},\ }\bibfield  {title} {\bibinfo {title} {Effective cahn-hilliard equation for the phase separation of active brownian particles},\ }\href {https://doi.org/10.1103/physrevlett.112.218304} {\bibfield  {journal} {\bibinfo  {journal} {Phys. Rev. Lett.}\ }\textbf {\bibinfo {volume} {112}},\ \bibinfo {pages} {218304} (\bibinfo {year} {2014})}\BibitemShut {NoStop}%
\bibitem [{\citenamefont {Nardini}\ \emph {et~al.}(2017)\citenamefont {Nardini}, \citenamefont {Fodor}, \citenamefont {Tjhung}, \citenamefont {van Wijland}, \citenamefont {Tailleur},\ and\ \citenamefont {Cates}}]{Nardini2017}%
  \BibitemOpen
  \bibfield  {author} {\bibinfo {author} {\bibfnamefont {C.}~\bibnamefont {Nardini}}, \bibinfo {author} {\bibfnamefont {Ã.}~\bibnamefont {Fodor}}, \bibinfo {author} {\bibfnamefont {E.}~\bibnamefont {Tjhung}}, \bibinfo {author} {\bibfnamefont {F.}~\bibnamefont {van Wijland}}, \bibinfo {author} {\bibfnamefont {J.}~\bibnamefont {Tailleur}},\ and\ \bibinfo {author} {\bibfnamefont {M.~E.}\ \bibnamefont {Cates}},\ }\bibfield  {title} {\bibinfo {title} {Entropy production in field theories without time-reversal symmetry: Quantifying the non-equilibrium character of active matter},\ }\href {https://doi.org/10.1103/physrevx.7.021007} {\bibfield  {journal} {\bibinfo  {journal} {Phys. Rev. X}\ }\textbf {\bibinfo {volume} {7}},\ \bibinfo {pages} {021007} (\bibinfo {year} {2017})}\BibitemShut {NoStop}%
\bibitem [{\citenamefont {Cahn}\ and\ \citenamefont {Hilliard}(1958)}]{Cahn1958}%
  \BibitemOpen
  \bibfield  {author} {\bibinfo {author} {\bibfnamefont {J.~W.}\ \bibnamefont {Cahn}}\ and\ \bibinfo {author} {\bibfnamefont {J.~E.}\ \bibnamefont {Hilliard}},\ }\bibfield  {title} {\bibinfo {title} {Free energy of a nonuniform system. i. interfacial free energy},\ }\href {https://doi.org/10.1063/1.1744102} {\bibfield  {journal} {\bibinfo  {journal} {J. Chem. Phys.}\ }\textbf {\bibinfo {volume} {28}},\ \bibinfo {pages} {258} (\bibinfo {year} {1958})}\BibitemShut {NoStop}%
\bibitem [{\citenamefont {Hohenberg}\ and\ \citenamefont {Halperin}(1977)}]{Hohenberg1977}%
  \BibitemOpen
  \bibfield  {author} {\bibinfo {author} {\bibfnamefont {P.~C.}\ \bibnamefont {Hohenberg}}\ and\ \bibinfo {author} {\bibfnamefont {B.~I.}\ \bibnamefont {Halperin}},\ }\bibfield  {title} {\bibinfo {title} {Theory of dynamic critical phenomena},\ }\href {https://doi.org/10.1103/revmodphys.49.435} {\bibfield  {journal} {\bibinfo  {journal} {Rev. Mod. Phys.}\ }\textbf {\bibinfo {volume} {49}},\ \bibinfo {pages} {435} (\bibinfo {year} {1977})}\BibitemShut {NoStop}%
\bibitem [{\citenamefont {Bray}(1994)}]{Bray1994}%
  \BibitemOpen
  \bibfield  {author} {\bibinfo {author} {\bibfnamefont {A.}~\bibnamefont {Bray}},\ }\bibfield  {title} {\bibinfo {title} {Theory of phase-ordering kinetics},\ }\href {https://doi.org/10.1080/00018739400101505} {\bibfield  {journal} {\bibinfo  {journal} {Adv. Phys.}\ }\textbf {\bibinfo {volume} {43}},\ \bibinfo {pages} {357} (\bibinfo {year} {1994})}\BibitemShut {NoStop}%
\bibitem [{\citenamefont {Theurkauff}\ \emph {et~al.}(2012)\citenamefont {Theurkauff}, \citenamefont {Cottin-Bizonne}, \citenamefont {Palacci}, \citenamefont {Ybert},\ and\ \citenamefont {Bocquet}}]{Theurkauff2012}%
  \BibitemOpen
  \bibfield  {author} {\bibinfo {author} {\bibfnamefont {I.}~\bibnamefont {Theurkauff}}, \bibinfo {author} {\bibfnamefont {C.}~\bibnamefont {Cottin-Bizonne}}, \bibinfo {author} {\bibfnamefont {J.}~\bibnamefont {Palacci}}, \bibinfo {author} {\bibfnamefont {C.}~\bibnamefont {Ybert}},\ and\ \bibinfo {author} {\bibfnamefont {L.}~\bibnamefont {Bocquet}},\ }\bibfield  {title} {\bibinfo {title} {Dynamic clustering in active colloidal suspensions with chemical signaling},\ }\href {https://doi.org/10.1103/PhysRevLett.108.268303} {\bibfield  {journal} {\bibinfo  {journal} {Phys. Rev. Lett.}\ }\textbf {\bibinfo {volume} {108}},\ \bibinfo {pages} {268303} (\bibinfo {year} {2012})}\BibitemShut {NoStop}%
\bibitem [{\citenamefont {Wittkowski}\ \emph {et~al.}(2014)\citenamefont {Wittkowski}, \citenamefont {Tiribocchi}, \citenamefont {Stenhammar}, \citenamefont {Allen}, \citenamefont {Marenduzzo},\ and\ \citenamefont {Cates}}]{Wittkowski2014}%
  \BibitemOpen
  \bibfield  {author} {\bibinfo {author} {\bibfnamefont {R.}~\bibnamefont {Wittkowski}}, \bibinfo {author} {\bibfnamefont {A.}~\bibnamefont {Tiribocchi}}, \bibinfo {author} {\bibfnamefont {J.}~\bibnamefont {Stenhammar}}, \bibinfo {author} {\bibfnamefont {R.~J.}\ \bibnamefont {Allen}}, \bibinfo {author} {\bibfnamefont {D.}~\bibnamefont {Marenduzzo}},\ and\ \bibinfo {author} {\bibfnamefont {M.~E.}\ \bibnamefont {Cates}},\ }\bibfield  {title} {\bibinfo {title} {Scalar $\phi^4$ field theory for active-particle phase separation},\ }\href {https://doi.org/10.1038/ncomms5351} {\bibfield  {journal} {\bibinfo  {journal} {Nat. Commun.}\ }\textbf {\bibinfo {volume} {5}},\ \bibinfo {pages} {4351} (\bibinfo {year} {2014})}\BibitemShut {NoStop}%
\bibitem [{\citenamefont {Tjhung}\ \emph {et~al.}(2018)\citenamefont {Tjhung}, \citenamefont {Nardini},\ and\ \citenamefont {Cates}}]{Tjhung2018}%
  \BibitemOpen
  \bibfield  {author} {\bibinfo {author} {\bibfnamefont {E.}~\bibnamefont {Tjhung}}, \bibinfo {author} {\bibfnamefont {C.}~\bibnamefont {Nardini}},\ and\ \bibinfo {author} {\bibfnamefont {M.~E.}\ \bibnamefont {Cates}},\ }\bibfield  {title} {\bibinfo {title} {Cluster phases and bubbly phase separation in active fluids: Reversal of the ostwald process},\ }\href {https://doi.org/10.1103/physrevx.8.031080} {\bibfield  {journal} {\bibinfo  {journal} {Phys. Rev. X}\ }\textbf {\bibinfo {volume} {8}},\ \bibinfo {pages} {031080} (\bibinfo {year} {2018})}\BibitemShut {NoStop}%
\bibitem [{\citenamefont {Fej\ifmmode~\mbox{\H{o}}\else \H{o}\fi{}s}\ \emph {et~al.}(2026)\citenamefont {Fej\ifmmode~\mbox{\H{o}}\else \H{o}\fi{}s}, \citenamefont {Sz\'ep},\ and\ \citenamefont {Yamamoto}}]{Fejos2025}%
  \BibitemOpen
  \bibfield  {author} {\bibinfo {author} {\bibfnamefont {G.}~\bibnamefont {Fej\ifmmode~\mbox{\H{o}}\else \H{o}\fi{}s}}, \bibinfo {author} {\bibfnamefont {Z.}~\bibnamefont {Sz\'ep}},\ and\ \bibinfo {author} {\bibfnamefont {N.}~\bibnamefont {Yamamoto}},\ }\bibfield  {title} {\bibinfo {title} {Scaling behaviors in active model $\mathrm{B}+$ via the functional renormalization group},\ }\href {https://doi.org/10.1103/k2qg-hrn1} {\bibfield  {journal} {\bibinfo  {journal} {Phys. Rev. E}\ }\textbf {\bibinfo {volume} {113}},\ \bibinfo {pages} {014130} (\bibinfo {year} {2026})}\BibitemShut {NoStop}%
\bibitem [{\citenamefont {Kardar}(2007)}]{Kardar2007}%
  \BibitemOpen
  \bibfield  {author} {\bibinfo {author} {\bibfnamefont {M.}~\bibnamefont {Kardar}},\ }\href@noop {} {\emph {\bibinfo {title} {Statistical Physics of Fields}}}\ (\bibinfo  {publisher} {Cambridge University Press},\ \bibinfo {address} {Cambridge},\ \bibinfo {year} {2007})\BibitemShut {NoStop}%
\bibitem [{\citenamefont {Goldenfeld}(2018)}]{Goldenfeld2018}%
  \BibitemOpen
  \bibfield  {author} {\bibinfo {author} {\bibfnamefont {N.}~\bibnamefont {Goldenfeld}},\ }\href {https://doi.org/10.1201/9780429493492} {\emph {\bibinfo {title} {Lectures on Phase Transitions and the Renormalization Group}}}\ (\bibinfo  {publisher} {CRC Press},\ \bibinfo {year} {2018})\BibitemShut {NoStop}%
\bibitem [{\citenamefont {Solon}\ \emph {et~al.}(2018{\natexlab{a}})\citenamefont {Solon}, \citenamefont {Stenhammar}, \citenamefont {Cates}, \citenamefont {Kafri},\ and\ \citenamefont {Tailleur}}]{Solon2018}%
  \BibitemOpen
  \bibfield  {author} {\bibinfo {author} {\bibfnamefont {A.~P.}\ \bibnamefont {Solon}}, \bibinfo {author} {\bibfnamefont {J.}~\bibnamefont {Stenhammar}}, \bibinfo {author} {\bibfnamefont {M.~E.}\ \bibnamefont {Cates}}, \bibinfo {author} {\bibfnamefont {Y.}~\bibnamefont {Kafri}},\ and\ \bibinfo {author} {\bibfnamefont {J.}~\bibnamefont {Tailleur}},\ }\bibfield  {title} {\bibinfo {title} {Generalized thermodynamics of phase equilibria in scalar active matter},\ }\href {https://doi.org/10.1103/PhysRevE.97.020602} {\bibfield  {journal} {\bibinfo  {journal} {Phys. Rev. E}\ }\textbf {\bibinfo {volume} {97}},\ \bibinfo {pages} {020602} (\bibinfo {year} {2018}{\natexlab{a}})}\BibitemShut {NoStop}%
\bibitem [{\citenamefont {T\"{a}uber}(2014)}]{Tuber2014}%
  \BibitemOpen
  \bibfield  {author} {\bibinfo {author} {\bibfnamefont {U.~C.}\ \bibnamefont {T\"{a}uber}},\ }\href {https://doi.org/10.1017/cbo9781139046213} {\emph {\bibinfo {title} {Critical Dynamics: A Field Theory Approach to Equilibrium and Non-Equilibrium Scaling Behavior}}}\ (\bibinfo  {publisher} {Cambridge University Press},\ \bibinfo {year} {2014})\BibitemShut {NoStop}%
\bibitem [{\citenamefont {Stanley}(1971)}]{Stanley1971}%
  \BibitemOpen
  \bibfield  {author} {\bibinfo {author} {\bibfnamefont {H.~E.}\ \bibnamefont {Stanley}},\ }\href {https://doi.org/10.1093/acprof:oso/9780195053166.001.0001} {\emph {\bibinfo {title} {Introduction to Phase Transitions and Critical Phenomena}}}\ (\bibinfo  {publisher} {Oxford University Press},\ \bibinfo {year} {1971})\BibitemShut {NoStop}%
\bibitem [{\citenamefont {Cardy}(1996)}]{Cardy1996}%
  \BibitemOpen
  \bibfield  {author} {\bibinfo {author} {\bibfnamefont {J.}~\bibnamefont {Cardy}},\ }\href {https://doi.org/10.1017/cbo9781316036440} {\emph {\bibinfo {title} {Scaling and Renormalization in Statistical Physics}}}\ (\bibinfo  {publisher} {Cambridge University Press},\ \bibinfo {year} {1996})\BibitemShut {NoStop}%
\bibitem [{\citenamefont {Privman}(1990)}]{Privman1990}%
  \BibitemOpen
  \bibinfo {editor} {\bibfnamefont {V.}~\bibnamefont {Privman}},\ ed.,\ \href {https://doi.org/10.1142/0988} {\emph {\bibinfo {title} {Finite Size Scaling and Numerical Simulation of Statistical Systems}}}\ (\bibinfo  {publisher} {World Scientific},\ \bibinfo {year} {1990})\BibitemShut {NoStop}%
\bibitem [{\citenamefont {Chaikin}\ and\ \citenamefont {Lubensky}(1995)}]{Chaikin1995}%
  \BibitemOpen
  \bibfield  {author} {\bibinfo {author} {\bibfnamefont {P.~M.}\ \bibnamefont {Chaikin}}\ and\ \bibinfo {author} {\bibfnamefont {T.~C.}\ \bibnamefont {Lubensky}},\ }\href {https://doi.org/10.1017/cbo9780511813467} {\emph {\bibinfo {title} {Principles of Condensed Matter Physics}}}\ (\bibinfo  {publisher} {Cambridge University Press},\ \bibinfo {year} {1995})\BibitemShut {NoStop}%
\bibitem [{\citenamefont {Ray}\ \emph {et~al.}(2024)\citenamefont {Ray}, \citenamefont {Mukherjee},\ and\ \citenamefont {Mohanty}}]{Ray2024}%
  \BibitemOpen
  \bibfield  {author} {\bibinfo {author} {\bibfnamefont {C.~G.}\ \bibnamefont {Ray}}, \bibinfo {author} {\bibfnamefont {I.}~\bibnamefont {Mukherjee}},\ and\ \bibinfo {author} {\bibfnamefont {P.~K.}\ \bibnamefont {Mohanty}},\ }\bibfield  {title} {\bibinfo {title} {How motility affects ising transitions},\ }\href {https://doi.org/10.1088/1742-5468/ad685b} {\bibfield  {journal} {\bibinfo  {journal} {J. Stat. Mech.: Theory Exp.}\ }\textbf {\bibinfo {volume} {2024}}\bibinfo  {number} { (9)},\ \bibinfo {pages} {093207}}\BibitemShut {NoStop}%
\bibitem [{\citenamefont {Landau}\ and\ \citenamefont {Binder}(2014)}]{LandauBinder2014}%
  \BibitemOpen
\bibfield  {number} {  }\bibfield  {author} {\bibinfo {author} {\bibfnamefont {D.~P.}\ \bibnamefont {Landau}}\ and\ \bibinfo {author} {\bibfnamefont {K.}~\bibnamefont {Binder}},\ }\href {https://doi.org/10.1017/CBO9781139696463} {\emph {\bibinfo {title} {A Guide to Monte Carlo Simulations in Statistical Physics}}},\ \bibinfo {edition} {4th}\ ed.\ (\bibinfo  {publisher} {Cambridge University Press},\ \bibinfo {year} {2014})\BibitemShut {NoStop}%
\bibitem [{\citenamefont {Partridge}\ and\ \citenamefont {Lee}(2019)}]{Partridge2019}%
  \BibitemOpen
  \bibfield  {author} {\bibinfo {author} {\bibfnamefont {B.}~\bibnamefont {Partridge}}\ and\ \bibinfo {author} {\bibfnamefont {C.~F.}\ \bibnamefont {Lee}},\ }\bibfield  {title} {\bibinfo {title} {Critical motility-induced phase separation belongs to the ising universality class},\ }\href {https://doi.org/10.1103/PhysRevLett.123.068002} {\bibfield  {journal} {\bibinfo  {journal} {Phys. Rev. Lett.}\ }\textbf {\bibinfo {volume} {123}},\ \bibinfo {pages} {068002} (\bibinfo {year} {2019})}\BibitemShut {NoStop}%
\bibitem [{\citenamefont {Solon}\ \emph {et~al.}(2018{\natexlab{b}})\citenamefont {Solon}, \citenamefont {Stenhammar}, \citenamefont {Cates}, \citenamefont {Kafri},\ and\ \citenamefont {Tailleur}}]{Solon2018IOP}%
  \BibitemOpen
  \bibfield  {author} {\bibinfo {author} {\bibfnamefont {A.~P.}\ \bibnamefont {Solon}}, \bibinfo {author} {\bibfnamefont {J.}~\bibnamefont {Stenhammar}}, \bibinfo {author} {\bibfnamefont {M.~E.}\ \bibnamefont {Cates}}, \bibinfo {author} {\bibfnamefont {Y.}~\bibnamefont {Kafri}},\ and\ \bibinfo {author} {\bibfnamefont {J.}~\bibnamefont {Tailleur}},\ }\bibfield  {title} {\bibinfo {title} {Generalized thermodynamics of motility-induced phase separation: phase equilibria, laplace pressure, and change of ensembles},\ }\href {https://doi.org/10.1088/1367-2630/aaccdd} {\bibfield  {journal} {\bibinfo  {journal} {New J. Phys.}\ }\textbf {\bibinfo {volume} {20}},\ \bibinfo {pages} {075001} (\bibinfo {year} {2018}{\natexlab{b}})}\BibitemShut {NoStop}%
\bibitem [{\citenamefont {Wilson}(1974)}]{Wilson1974}%
  \BibitemOpen
  \bibfield  {author} {\bibinfo {author} {\bibfnamefont {K.}~\bibnamefont {Wilson}},\ }\bibfield  {title} {\bibinfo {title} {The renormalization group and the ε expansion},\ }\href {https://doi.org/10.1016/0370-1573(74)90023-4} {\bibfield  {journal} {\bibinfo  {journal} {Phys. Rep.}\ }\textbf {\bibinfo {volume} {12}},\ \bibinfo {pages} {75} (\bibinfo {year} {1974})}\BibitemShut {NoStop}%
\bibitem [{\citenamefont {Kosterlitz}\ and\ \citenamefont {Thouless}(1973)}]{Kosterlitz1973}%
  \BibitemOpen
  \bibfield  {author} {\bibinfo {author} {\bibfnamefont {J.~M.}\ \bibnamefont {Kosterlitz}}\ and\ \bibinfo {author} {\bibfnamefont {D.~J.}\ \bibnamefont {Thouless}},\ }\bibfield  {title} {\bibinfo {title} {Ordering, metastability and phase transitions in two-dimensional systems},\ }\href {https://doi.org/10.1088/0022-3719/6/7/010} {\bibfield  {journal} {\bibinfo  {journal} {J. Phys. C: Solid State Phys.}\ }\textbf {\bibinfo {volume} {6}},\ \bibinfo {pages} {1181} (\bibinfo {year} {1973})}\BibitemShut {NoStop}%
\bibitem [{\citenamefont {Bray}(2002)}]{Bray2002}%
  \BibitemOpen
  \bibfield  {author} {\bibinfo {author} {\bibfnamefont {A.~J.}\ \bibnamefont {Bray}},\ }\bibfield  {title} {\bibinfo {title} {Theory of phase-ordering kinetics},\ }\href {https://doi.org/10.1080/00018730110117433} {\bibfield  {journal} {\bibinfo  {journal} {Adv. Phys.}\ }\textbf {\bibinfo {volume} {51}},\ \bibinfo {pages} {481} (\bibinfo {year} {2002})}\BibitemShut {NoStop}%
\bibitem [{\citenamefont {Cates}\ and\ \citenamefont {Tjhung}(2017)}]{Cates2017}%
  \BibitemOpen
  \bibfield  {author} {\bibinfo {author} {\bibfnamefont {M.~E.}\ \bibnamefont {Cates}}\ and\ \bibinfo {author} {\bibfnamefont {E.}~\bibnamefont {Tjhung}},\ }\bibfield  {title} {\bibinfo {title} {Theories of binary fluid mixtures: from phase-separation kinetics to active emulsions},\ }\href {https://doi.org/10.1017/jfm.2017.832} {\bibfield  {journal} {\bibinfo  {journal} {J. Fluid Mech.}\ }\textbf {\bibinfo {volume} {836}},\ \bibinfo {pages} {P1} (\bibinfo {year} {2017})}\BibitemShut {NoStop}%
\bibitem [{\citenamefont {Pattanayak}\ \emph {et~al.}(2021)\citenamefont {Pattanayak}, \citenamefont {Mishra},\ and\ \citenamefont {Puri}}]{Pattanayak2021}%
  \BibitemOpen
  \bibfield  {author} {\bibinfo {author} {\bibfnamefont {S.}~\bibnamefont {Pattanayak}}, \bibinfo {author} {\bibfnamefont {S.}~\bibnamefont {Mishra}},\ and\ \bibinfo {author} {\bibfnamefont {S.}~\bibnamefont {Puri}},\ }\bibfield  {title} {\bibinfo {title} {Ordering kinetics in the active model $\mathrm{B}$},\ }\href {https://doi.org/10.1103/PhysRevE.104.014606} {\bibfield  {journal} {\bibinfo  {journal} {Phys. Rev. E}\ }\textbf {\bibinfo {volume} {104}},\ \bibinfo {pages} {014606} (\bibinfo {year} {2021})}\BibitemShut {NoStop}%
\bibitem [{\citenamefont {Yadav}\ \emph {et~al.}(2025)\citenamefont {Yadav}, \citenamefont {Mishra},\ and\ \citenamefont {Puri}}]{Yadav2025}%
  \BibitemOpen
  \bibfield  {author} {\bibinfo {author} {\bibfnamefont {P.~K.}\ \bibnamefont {Yadav}}, \bibinfo {author} {\bibfnamefont {S.}~\bibnamefont {Mishra}},\ and\ \bibinfo {author} {\bibfnamefont {S.}~\bibnamefont {Puri}},\ }\bibfield  {title} {\bibinfo {title} {Coarsening kinetics in active model $\mathrm{B}+$: Macroscale and microscale phase separation},\ }\href {https://doi.org/10.1103/kb9k-w7jr} {\bibfield  {journal} {\bibinfo  {journal} {Phys. Rev. E}\ }\textbf {\bibinfo {volume} {112}},\ \bibinfo {pages} {035412} (\bibinfo {year} {2025})}\BibitemShut {NoStop}%
\end{thebibliography}%

\end{document}